\documentclass[aps, prl, 10pt, twocolumn, superscriptaddress, amsmath, amssymb]{revtex4-2}

\usepackage[T1]{fontenc} % for names in bibliography with exotic letters

\usepackage{graphicx}
\usepackage{upgreek}
\usepackage{natbib}
\usepackage{physics}
\usepackage{siunitx}
\usepackage{bbold}
\usepackage{bm} % bold math
\usepackage{enumitem}
\setlist{nosep} % part of enumitem - no separation between lists and items in lists
\usepackage{textcomp}
\usepackage[dvipsnames]{xcolor} % for coloring the text
\usepackage[colorlinks=true,citecolor=blue,linkcolor=blue]{hyperref}

\newcommand{\NVm}[0]{NV$^{-}$}

\begin{document}

\title{Temperature Dependent Photophysics of Single NV Centers in Diamond}

\author{Jodok Happacher}
\affiliation{Department of Physics, University of Basel, Klingelbergstrasse 82, Basel CH-4056, Switzerland}
\author{Juanita Bocquel}
\affiliation{Department of Physics, University of Basel, Klingelbergstrasse 82, Basel CH-4056, Switzerland}
\author{Hossein T. Dinani}
\affiliation{Escuela Data Science, Facultad de Ciencias, Ingeniería y Tecnología, Universidad Mayor, Santiago, Chile}
\author{M\"arta A. Tschudin}
\affiliation{Department of Physics, University of Basel, Klingelbergstrasse 82, Basel CH-4056, Switzerland}
\author{Patrick Reiser}
\affiliation{Department of Physics, University of Basel, Klingelbergstrasse 82, Basel CH-4056, Switzerland}
\author{David A. Broadway}
\affiliation{Department of Physics, University of Basel, Klingelbergstrasse 82, Basel CH-4056, Switzerland}
\author{Jeronimo R. Maze}
\affiliation{Facultad de F\'isica, Pontificia Universidad Cat\'olica de Chile, Santiago 7820436, Chile}
\author{Patrick Maletinsky}
\email[]{patrick.maletinksy@unibas.ch}
\affiliation{Department of Physics, University of Basel, Klingelbergstrasse 82, Basel CH-4056, Switzerland}

\date{\today}

\begin{abstract}
We present a comprehensive study of the temperature and magnetic-field dependent photoluminescence (PL) of individual NV centers in diamond, spanning the temperature-range from cryogenic to ambient conditions.
We directly observe the emergence of the NV's room-temperature effective excited state structure and provide a clear explanation for a previously poorly understood broad quenching of NV PL at intermediate temperatures around $50~$K. 
We develop a model that quantitatively explains all of our findings, including the strong impact that strain has on the temperature-dependence of the NV's PL. 
These results complete our understanding of orbital averaging in the NV excited state and have significant implications for the fundamental understanding of the NV center and its applications in quantum sensing.
\end{abstract}
\maketitle

Color centers in solid state hosts are crucial for a variety of quantum technologies, including spin-based quantum sensors\,\cite{Degen2017a}, highly stable fluorescent labels\,\cite{Alkahtani2018a}, and single-photon light sources for advanced microscopy\,\cite{Nelz2020a}. 
Among the many potential systems, the nitrogen vacancy (NV) lattice defect in diamond stands out due to its multiple demonstrated applications in areas such as nanoscale imaging\,\cite{Thiel2016a,Thiel2019a} and quantum information processing\,\cite{Hensen2015a,Kalb2017}, as well as its robustness in a wide range of environmental conditions\,\cite{Toyli2012a,Fu2020,Lesik2019a}, including promising use-cases of nanoscale magnetometry in cryogenic conditions\,\cite{Thiel2016a,Pelliccione2016a,Thiel2019a}.

\begin{figure}[th!]
	\includegraphics[width=1\linewidth]{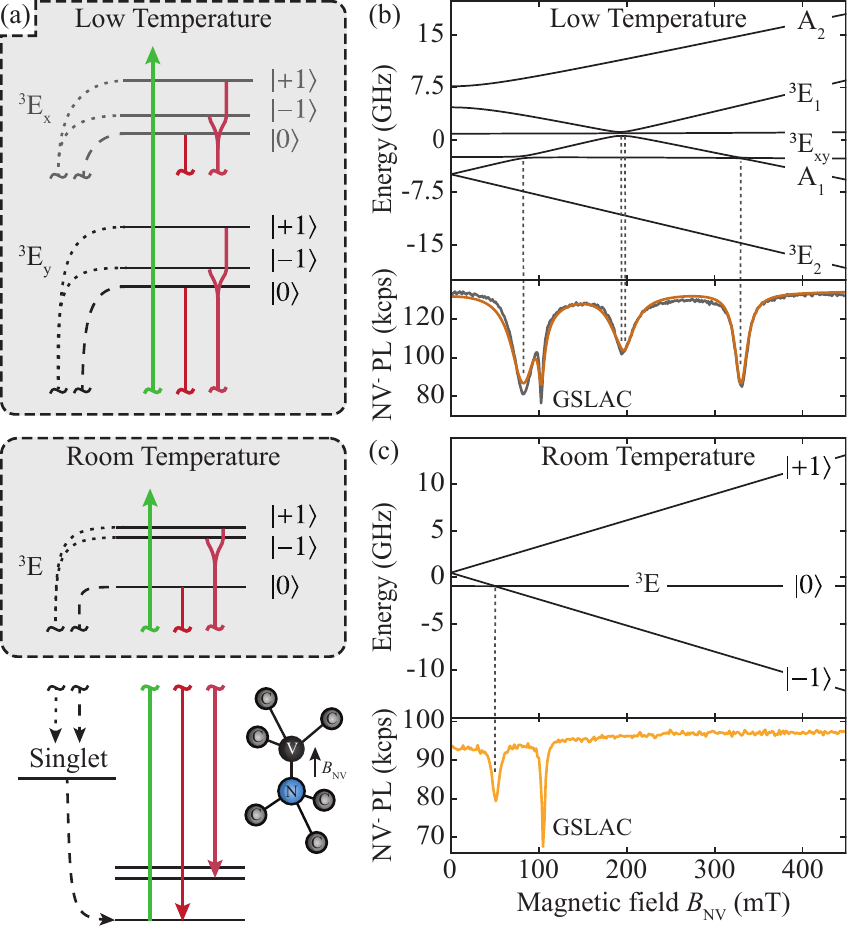}
	\caption{
		(a)~NV level structure for optical spin pumping and spin readout with the excited state manifold for both low temperatures (top panel) and the orbital averaged room temperature (middle panel). Additionally, the spin singlet and ground state is also shown which is applicable for both regimes (bottom panel). 
		(b)~The excited state manifold for an NV spin at $T = 2~$K with relatively low strain $\delta_\perp = 1.683 \pm 0.003$~GHz (top panel) and the corresponding NV PL as a function of applied magnetic field. 
		(c)~Same as in (b) but for $T = 300~$K, for the same low strain NV. 
	}
	\label{Fig:NVLevelStruct}
\end{figure}

%Principle spin readout and init
Most applications of NV centers rely on their highly coherent ground state electron spin\,\cite{Balasubramanian2009a} and the ability to efficiently initialize\,\cite{Harrison2004, Tetienne2012} and read out\,\cite{Dreau2011,Steiner2010} the spin optically. 
These techniques are based on a spin-dependent intersystem crossing from the NV's optical excited state to a metastable spin-singlet manifold, from which the system decays into a well-defined spin state of the NV's ground state (GS)\,\cite{Doherty2013a} (Fig.\,\ref{Fig:NVLevelStruct}a).

%%FIGURE: NV Level schemes and basic PL vs B data at LT and RT

%ES - LT vs RT 
This intersystem crossing, and therefore the mechanism of NV spin readout and initialisation, results from the properties of the NV's orbital excited states (ES) and their coupling to the NV's $^1$A$_1$ singlet state\,\cite{Doherty2013a}.
It is remarkable that, while spin initialisation and readout are observed both at cryogenic and ambient conditions, the effective ES level structures are markedly different in the two cases. 
At temperatures below few tens of Kelvins, and in the limit of sizeable strain, the NV ES exhibits two orbital branches, commonly denoted as $E_x$ and $E_y$ (Fig.\,\ref{Fig:NVLevelStruct}a)\,\cite{Doherty2013a}.
Each of these branches in turn splits into three electronic spin sub-levels with magnetic quantum numbers $m_s=-1,0,+1$. 
Conversely, at temperatures $T\gtrsim100~$K,  phonon induced orbital averaging\,\cite{Fu2009a} effectively reduces the NV ES to a single orbital with spin~$1$ where states of magnetic quantum numbers $m_s=\pm1$ are split from the $m_s=0$ state by a zero-field splitting of $D_0^{\rm ES}/h=1.4~$GHz   (Fig.\,\ref{Fig:NVLevelStruct}a)\,\cite{Fu2009a,Batalov2009a,Fuchs2008a}.

%%FIGURE: I vs B - model and simulations
\begin{figure*}[t]
	\includegraphics[width=1\linewidth]{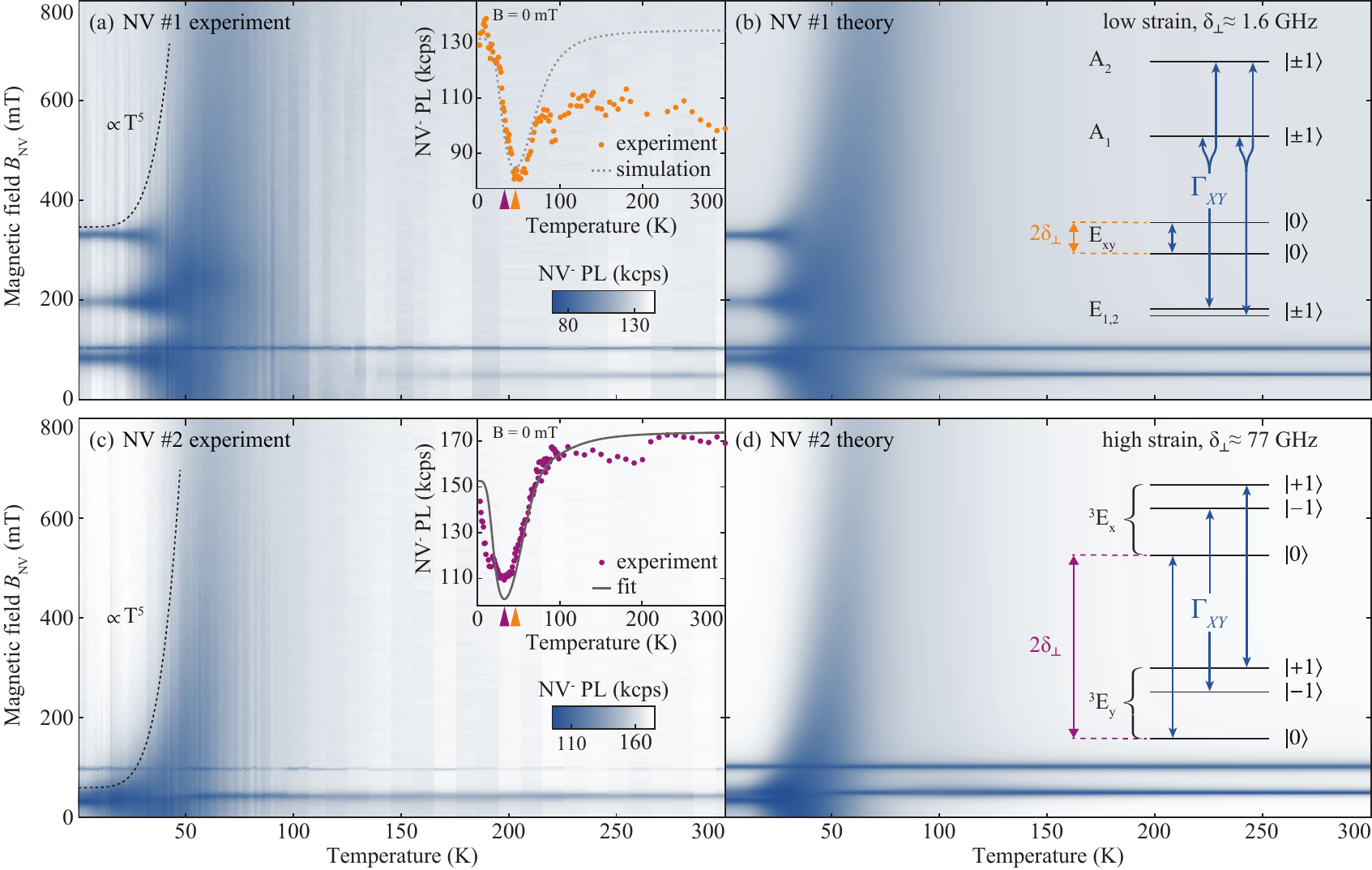}
	\caption{
		(a)~Experimental data of photoluminescence (PL) intensity $I_{\rm PL}$ recorded on a single NV center with ``low'' strain (strain parameter $\delta_\perp=1.683 \pm 0.003~$GHz) as a function of temperature $T$ and magnetic field $B_{\rm NV}$ applied along the NV symmetry axis.
		$I_{\rm PL}$ data are normalized as described in the text.
		The dotted line is a guide to the eye for which $B_{\rm NV}\propto T^5$, scaled to follow the half-width contour of the $I_{\rm PL}$ dip. 
		Inset: Raw $I_{\rm PL}(T)$ data obtained at $B_{\rm NV}=0~$mT, together with a model simulation using the extracted phonon coupling from NV \#2 data .
		(b)~Model prediction of $I_{\rm PL}(B_{\rm NV},T)$ (see text), with relevant NV excited state level structure (inset). The rate $\Gamma_{XY}$ is extracted from fits to NV \#2 data as illustrated in the inset to panel (c). All other NV transition rates are taken from literature\,\cite{Goldman2015b}. 
		(c)~Experimental data as in panel (a), but here taken on a single NV center with ``high'' strain (strain parameter $\delta_\perp=77\pm3~$GHz).
		Inset: Raw $I_{\rm PL}(T)$ data obtained at $B_{\rm NV}=0~$mT, together with the fit of the model to the raw data.
		(d)~Model prediction as in panel (b) here for the ``high'' strain case. 
		All other rates in the model remain identical.
	}
	\label{fig:PLvsBvsT}
\end{figure*}

Interestingly, while the foundations of intersystem crossing\,\cite{Goldman2015a,Goldman2015b}  and orbital averaging\,\cite{Fu2009a,Rogers2009a,Plakhotnik2015a} have been studied in the past, the transition between the low-temperature and the high-temperature limits, and the emergence of the RT ES spin structure, have never been explored in a systematic way. 
Prior work on NV ensembles has established a non-trivial temperature dependence of the NV PL intensity at zero magnetic field\,\cite{Rogers2009a}, including a local minimum of the NV PL at $T\approx40~$K, which remains unexplained thus far. 
In addition, it was experimentally shown through optical linewidth measurements that orbital averaging between $E_x$ and $E_y$ is dominated by two-phonon mixing processes whose rate scales with temperature as $T^5$\,\cite{Fu2009a}. 
So far, however, it remained unexplained how such orbital averaging can account for the non-trivial temperature dependence of the NV's PL intensity, and how the established, room-temperature behaviour of the NV ES emerges from this picture.

Here, we present a systematic experimental study of the NV photoluminescence (PL) intensity, $I_{\rm PL}$, as a function of both magnetic field and temperature in the range $T=2-300~$K, that offers a concise and complete picture of the NV's temperature-dependant photophysics. 
Our work builds on recent results\,\cite{Happacher2022a} that revealed key fingerprints of the NV's cryogenic ES level structure through dips in $I_{\rm PL}$ that occur at specific magnetic fields $B_{\rm NV}$ applied along the NV quantisation axis.
These dips are the result of ES level anti-crossings (ESLACs) between levels of unlike $m_s$ that cause an intersystem crossing into the dark singlet states and therefore a drop in $I_{\rm PL}$.
This process is illustrated by the data presented in Fig.\,\ref{Fig:NVLevelStruct}b that shows measurements of $I_{\rm PL}$ as a function of $B_{\rm NV}$ at $T=2~$K, for an NV with a relatively low strain-induced $E_x$-$E_y$ level splitting $\delta_\perp\approx1.6~$GHz\,\cite{Happacher2022a}.
In this work, we exploit this approach to explore the transition of the ES level structure from cryostat base temperature ($T\approx 2~$K) to ambient ($T\approx 300~$K) and track the emergence to the resulting high-temperature ES level structure (Fig.\,\ref{Fig:NVLevelStruct}c).

%Samples and experimental setting
Our experiments were performed on two representative, single NV centers that we studied in a variable-temperature, cryogenic confocal microscope (see\,\cite{SOM} for details). 
The NV centers were located within few microns of the surface of two ultra-pure diamond samples (NV\,$\#1$: $(100)-$oriented, electronic grade, Element6; NV\,$\#2$: $(111)-$oriented, custom-grown diamond\,\cite{Neu2014a}), and were each embedded in diamond photonic structures to enhance optical collection efficiency\,\cite{Jamali2014a}.
The two NVs differed in the magnitude of the ES strain splitting parameter $\delta_\perp$, which for NV\,$\#1$ was comparable ($\delta_\perp=1.683 \pm 0.003~$GHz) to, and for NV\,$\#2$ much larger ($\delta_\perp=77 \pm 3~$GHz) than the NV ES fine-structure splittings, which are $\lesssim 5~$GHz\,\cite{Doherty2013a}.
The quoted strain values were extracted through a model-fit to $I_{\rm PL}(B_{\rm NV})$ at $T=2~$K, as described elsewhere\,\cite{Happacher2022a}.  
In all cases, we recorded $I_{\rm PL}$ while sweeping $B_{\rm NV}$ and stepping $T$ under conditions of moderate optical excitation (excitation intensity $\approx 1 \times$ optical saturation) with green laser light ($\lambda=532~$nm).

To mitigate the effect of small but unavoidable signal drifts during our experiments, we normalized our data. 
For this, we performed a fit of the model (see Eq.~\ref{Eq:LME} and following) to the raw $I_{\rm PL}(T)$ data of NV $\# 2$.
The fit shows qualitatively good agreement to the raw data (e.g. at $B=0$~mT Fig.\,\ref{fig:PLvsBvsT}c, inset), with deviations occurring at temperatures where signal drifts were dominant.
We therefore normalized each dataset $I_{\rm PL}(B_{\rm NV})$ to the expected value at $B_{\rm NV}=800~$mT to arrive at the final dataset\,\footnote{Non-normalized raw data are shown in the supplementary information\,\cite{SOM} and shows qualitative agreement with the data shown in Fig.\,\ref{fig:PLvsBvsT}a and c.}.

%Experimental data - PL vs B vs T; high strain
These normalized $I_{\rm PL}$ data as a function of $B_{\rm NV}$ and $T$, taken for the low-strain NV\,$\#1$ are shown in Fig.\,\ref{fig:PLvsBvsT}a.
At $T=2~$K, the NV level structure of NV\,$\#1$ (Fig.\,\ref{Fig:NVLevelStruct}b) results in four sharp $I_{\rm PL}$ dips, which arise from ESLACs occurring both within and between the orbital branches $E_x$ and $E_y$\,\cite{Happacher2022a}.
With increasing temperature, these ESLAC dips start to broaden, and reach a maximal width at $T\approx60~$K, where they span almost the entire magnetic field range accessible in our experiment. 
Qualitatively, this broadening is well-described by the $T^5$ scaling expected from two-phonon orbital mixing processes\,\cite{Fu2009a} (dotted line in Fig.\,\ref{fig:PLvsBvsT}a). 

Remarkably, upon further increasing the temperature, the strongly broadened ESLAC dips disappear and between $T\approx70-150~$K, the only discernible feature in $I_{\rm PL}(B_{\rm NV})$ is a narrow dip located at $B_{\rm NV}=102.5~$mT.
This dip results from the NV's well-understood GS level anti-crossing (GSLAC)\,\cite{Broadway2016a, Wickenbrock2016a}. 
Only at significantly higher temperatures $T\approx150~$K, the single, sharp dip at $B_{\rm NV}=50.5~$mT appears, that corresponds to the NV's well-known RT ESLAC corresponding to the level structure illustrated in Fig.\,\ref{Fig:NVLevelStruct}c.

%Experimental data - PL vs B vs T; high strain
We repeated our experiment on the high-strain NV\,$\#2$ (Fig.\,\ref{fig:PLvsBvsT}\,c). 
Compared to our findings on NV\,$\#1$, we find several differences in the evolution of $I_{\rm PL}(B_{\rm NV})$ with temperature.  
At $T=2~$K, the high-strain NV level structure of NV\,$\#2$ now results in only two $I_{\rm PL}$ dips, each corresponding to ESLACs occurring within one of the orbital branches\,\cite{Happacher2022a}.
These two ESLAC dips again show a broadening $\propto T^5$ (dotted line in Fig.\,\ref{fig:PLvsBvsT}c) and merge to a single, broad $I_{\rm PL}$ dip covering the whole accessible range in $B_{\rm NV}$ for $T\approx 50~$K.
Upon further increasing the temperature, the strongly broadened ESLAC dips again disappear and, starting from $T\approx60~$K, are replaced by the RT ESLAC dip discussed before.
The simulation uses the magnetic field misalignment to the NV axis measured at low temperature.
Some differences between model and data over the entire temperature range can be explained by small variations of this alignment.

%Model
To obtain a full understanding of our data, we developed a quantitative model to describe the NV population dynamics as a function of $B_{\rm NV}$ and $T$, that explicitly takes into account the full, low-temperature ES NV level structure\,\cite{Maze2011a} and temperature-dependant, phonon-induced relaxation processes between the involved ESs. 
For this, we determine the dynamics of the NV's density matrix $\hat{\rho}$ through the Lindblad master equation
\begin{equation}
\label{Eq:LME}
\frac{d}{dt}\hat{\rho}=\frac{i}{\hbar}[\hat{\rho},\hat{H}_{\rm NV}]+\sum_k\hat{L}_k\hat{\rho}\hat{L}_k^{\dagger}-\frac{1}{2}\{\hat{L}_k^{\dagger}\hat{L}_k,\hat{\rho}\},
\end{equation}
where $\hat{H}_{\rm NV}$, and $\hat{L}_k$ are the NV Hamiltonian and the relevant collapse operators, respectively. 
The $10\times10$ density matrix $\hat{\rho}$ comprises the level structure depicted in Fig.\,\ref{Fig:NVLevelStruct}a, specifically the three $^3$A$_2$ GSs, the six $^3$E states, and a single state representing the singlet shelving states $^1$A$_1$ and E$_1$. 
For $\hat{H}_{\rm NV}$, we employ the well-established NV Hamiltonians for the $^3$A$_2$ and $^3$E manifolds\,\cite{Doherty2013a,Maze2011a}.
Explicit definitions and expression for $\hat{\rho}$ and $\hat{H}_{\rm NV}$ are given in\,\cite{SOM}. 

%Relaxation and orbital averaging
The collapse operators 
$\hat{L}_{\ket{i} \to \ket{j}}=\sqrt{\Gamma_{ij}}\ket{j}\bra{i}$ 
describe transitions from states $\ket{i}$ to $\ket{j}$ occurring at rates $\Gamma_{ij}$.
The temperature dependence of our $I_{\rm PL}$ data is fully explained by considering the spin preserving, phonon-induced couplings between the orbital branches $E_x$ and $E_y$\,\cite{Fu2009a}, that are described by $\hat{L}_{\ket{X} \to \ket{Y}}=\sqrt{\Gamma_{XY}(T)}\ket{Y}\bra{X}$ and its inverse process.\,\footnote{We consider symmetric rates, i.e. $\Gamma_{XY}=\Gamma_{YX}$, only, since the slight asymmetry caused by spontaneous decay $\ket{X} \to \ket{Y}$ has no measurable effect on our results. The reason for this is that the asymmetry only becomes noticeable once $\Gamma_{XY}\ll\gamma_{\rm NV}$, in which case the ES dynamics is governed by spontaneous decay into the NV GS.}.
Here, $\ket{X}$ ($\ket{Y}$) are state vectors corresponding to the ES $E_x$ ($E_y$) manifolds.
In addition, we model optical excitation, spontaneous emission and intersystem crossings with collapse operators between the corresponding orbital manifolds, at constant rates that we obtain from literature\,\cite{Gupta2016a,Goldman2015b} and keep constant\,\footnote{We note that our main findings are insensitive to the rates employed and we find largely identical results by employing other literature values.}.
We note that temperature dependencies have only been reported for intersystem crossing rates for $T<20$~K\,\cite{Goldman2015a}, and in this regime have a negligible effect to our findings\,\cite{Happacher2022a}. 
The explicit expressions for all collapse operators we employ are given in\,\cite{SOM}.

%Discussion of modelling results
To model the behaviour of $I_{\rm PL}(B_{\rm NV},T,\delta_\perp)$, we numerically solve for the steady state solution of Eq.\,(\ref{Eq:LME}) and determine $I_{\rm PL}$ as being proportional to the total NV ES population for each value of $B_{\rm NV}$ and $T$ (Figs.\,\ref{fig:PLvsBvsT}\,b and\,d).
The model shows remarkable qualitative and quantitative agreement with our data, both for the high-strain and low-strain cases. 
The emergence of the RT-ESLAC is quantitatively reproduced in both cases, including the marked difference of the onset temperature for the RT-ESLAC that we discuss below. 

Our modelling results offer an intuitive way to understand our experimental findings and the evolution of the ESLAC induced $I_{\rm PL}$ dips with temperature.
At low temperatures, $T\lesssim10~$K, $I_{\rm PL}(B_{\rm NV})$ is unaffected by orbital averaging and only starts to broaden once $\Gamma_{XY}\approx\gamma_{\rm NV}$, where $\gamma_{\rm NV}=(12.5~{\rm ns})^{-1}$ is the NV's radiative recombination rate.
With increasing temperature, the ES levels and therefore the ESLACs start to broaden due to orbital mixing at the rate $\Gamma_{XY}$, i.e. $\propto T^5$.
The recovery of $I_{\rm PL}(B_{\rm NV})$ at $T\gtrsim60~$K can then be understood as a process akin to motional narrowing in nuclear magnetic resonance\,\cite{Slichter2013a}:
Once $\Gamma_{XY}$ exceeds $\lambda_{es}^\perp$ -- the rate of spin-mixing in the NV's ES\,\cite{SOM,Happacher2022a}) -- 
jumps between orbital states will interrupt, and therefore effectively suppress ES spin mixing processes.
Since the reduction of $I_{\rm PL}$ results from ESLAC induced spin-mixing and subsequent shelving into the NV singlet states, $I_{\rm PL}$ will recover once $\Gamma_{XY}\gtrsim\lambda_{es}^\perp$.
After this point in temperature, $I_{\rm PL}(B_{\rm NV})$ is governed by the effective, RT ES level structure (Fig.\,\ref{Fig:NVLevelStruct}\,c) and shows the well-known RT-ESLAC.
This regime is well described by our model but can alternatively also be derived by taking the partial trace of $\hat{H}_{\rm NV}$ over the orbital degrees of freedom\,\cite{Plakhotnik2014a,SOM}.

%Discussion phonon coupling fit results
We note that the quantitative nature of our model also allows us to determine the phonon-induced mixing rate $\Gamma_{XY}$ from fits to the data such as the one presented in Fig.\,\ref{fig:PLvsBvsT}c, inset.
For NV $\# 1$ it yields 
$\Gamma_{XY}=(1+\epsilon(T,\delta_{\perp}))\cdot\gamma_{\rm NV}\cdot(1.10 \pm 0.05 \cdot 10^{-6}K^{-5}) \times T^5$,
where $\epsilon$ is a weakly temperature- and strain-dependent correction-factor, with $\epsilon\ll1$ for T$\approx10\dots100~$K\,\cite{SOM}.
This rate agrees well with prior results\,\cite{Fu2009a,Goldman2015a,Plakhotnik2015a}, but was here obtained in a complementary way that does not require complex, resonant laser spectroscopy.

\begin{figure}
	\includegraphics[width=1\linewidth]{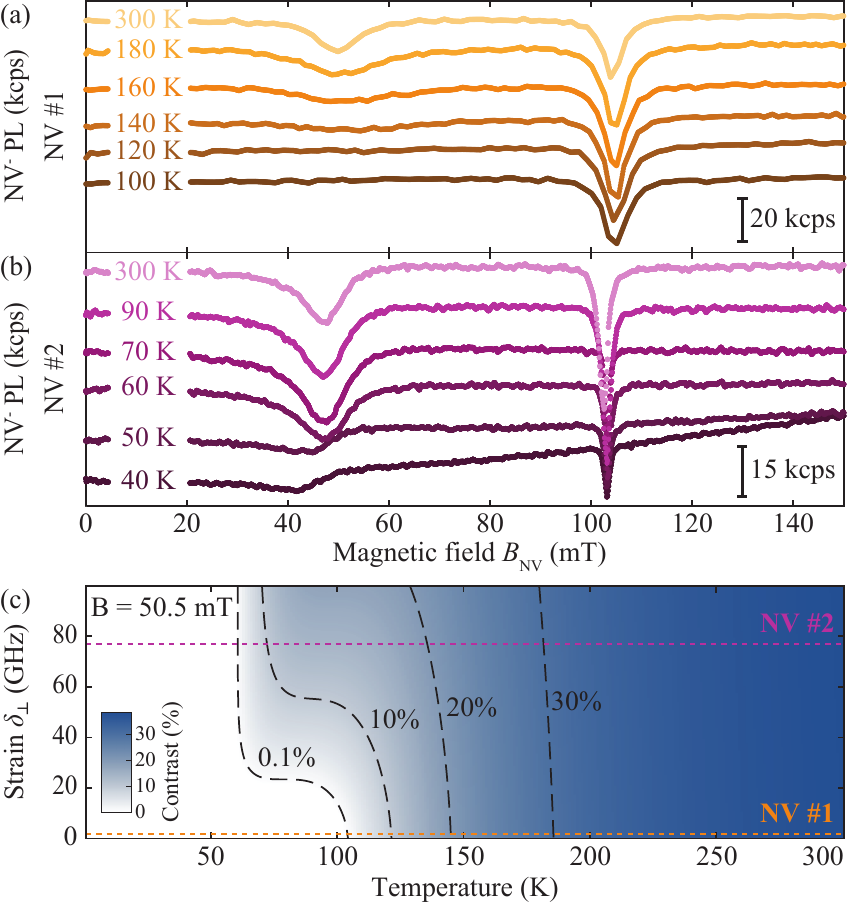}
	
	\caption{
		Appearance of the room-temperature excited state level anti-crossing (RT-ESLAC) for the NVs with low strain (a) and high strain (b). 
		The data is offset for clarity.
		For the low-strain NV, the RT-ESLAC appears only at significantly higher temperature compared to the high-strain NV.
		(c) Model prediction of the relative change in $I_{\rm PL}(T,\delta_\perp)$, evaluated at the RT-ESLAC field (see text and\,\cite{SOM}).}
	\label{fig:PLvsB_T_RT_ESLAC}
\end{figure}

Lastly, we note that our model and data also yield the unexpected observation that the appearance of the RT-ESLAC has a strong strain dependence.
The data in Fig.\,\ref{fig:PLvsB_T_RT_ESLAC}a and b illustrates this and shows how the RT-ESLAC appears much later for the low-strain NV\,\#1, as compared to the high-strain NV\,\#2.
To further support this observation, we extracted from our model the contribution to $I_{\rm PL}$ originating from the RT-ESLAC alone\,\cite{SOM}.
We present the resulting model prediction for the relative change in $I_{\rm PL}(T,\delta_\perp)$, evaluated at $B_{\rm NV}=50.5~$mT, i.e. at the ESLAC field, in Fig.\,\ref{fig:PLvsB_T_RT_ESLAC}c. 
The simulation clearly evidences the strong strain-dependence of the onset temperature for the RT-ESLAC, which qualitatively reproduces the experimental data. 
While the temperature dependence of the ES ODMR for low-strain NVs has been assessed in a prior study\,\cite{Batalov2009a}, its quantitative understanding that we present here has been missing thus far.

%%FIGURE: Appearance of the RT-ESLAC

%Conclusion, Summary and outlook
In conclusion, we have presented a comprehensive study of the magnetic field and temperature dependence of the PL emission rates of individual NV centers in the limit of both low and high strain.
Our work presents a complete picture and a quantitative model of the temperature-induced orbital averaging process that was missing thus far. 
It thereby complements past research on orbital averaging in the NV ES and allows for deeper insight into the orbital averaging and the emergence of an effective, room-temperature ES level structure.
Next to fundamental insights into the spectroscopic properties of NV centers, our results are of relevance to applications of NV centers in quantum sensing and quantum information processing, in that they predict allowed regions of operation in the parameter-space of magnetic field and temperature. 
Specifically, our experimental data and accompanying theory allow one to identify operational conditions away from ESLACs, where optical spin-initialisation and readout is most effective. 

Importantly, our results have implications beyond pure NV center based research.
In particular, the methods presented here apply to any color-center where ESLACs and spin-dependent dark states (i.e. optically detected spin resonance) occurs. 
Examples for this include SiV centers in SiC\,\cite{Wang2020a} or the charge-neutral SiV center in diamond\,\cite{Rose2018a}, where our method could shed new light into unknown ES structures, orbital averaging, or still poorly understood temperature dependencies in SiV$^0$ PL\,\cite{Haenens-Johansson2011a}.

{\em Note added:} During completion of this work, we became aware of related work on NV PL studied at selected temperatures and magnetic fields\,\cite{Ernst2023a}.

%% Acknowledgements
We gratefully acknowledge N.~Manson, T.~F.~Sjolander and V.~Jacques for fruitful discussions, and A.~Tallaire and J.~Achard for growth of the sample containing NV\,\#2.
We further acknowledge financial support from ANID-Fondecyt $1221512$ and ANID-Anillo ACT $192023$, from NCCR QSIT (Grant No. $185902$), the Swiss Nanoscience Institute, and through the Swiss NSF Project Grant No. $188521$.
%%%%%%%%%%%%%%%%%%%%%%

\bibliographystyle{apsrev4-2}
\bibliography{Bibliography_NV_Temperature_Dependence}

\clearpage

\onecolumngrid
\appendix

\section*{Supplementary Information}

\tableofcontents

\section{Theoretical description}

In order to model the temperature dependence of the photo-physics of single NV centers we use a Lindbladian master equation to calculate the time evolution of the population in a 10-level system, similar to the work presented in previous work in Ref.~\cite{Happacher2022a}. 
This formalism facilitates the inclusion of temperature dependent effects, which are caused by electron-phonon interactions, directly into the low temperature \NVm{} Hamiltonian, extending this model to a larger temperature range. 
Using this Hamiltonian, the NV spin's photoluminescence (PL) is subsequently calculated from the resulting steady-state populations.

\subsection{NV spin Hamiltonians}\label{Subsection:H}

For an appropriate description of the NV photo-dynamics at low temperatures, both the ground and excited states Hamiltonians need to be considered. 
Here we construct a model using the Hamiltonian previously described by Doherty et al\,\cite{Doherty2013a}, which is equivalent to using the alternative Hamiltonian form described by Maze et al\,\cite{Maze2011a}.

The canonical spin-Hamiltonian of the NV spin's ground state is
\begin{equation}
\hat{\mathcal{H}}_{gs}=D_{gs}\Big[\hat{S}_{z}^{2}-S(S+1)/3~\mathbb{1}_{3}\Big] ,
\end{equation}
where $D_{gs} \approx 2.88$ GHz, $S = 1$ for a spin 1 system, $\hat{S}_{z}$ is the spin operator and $\mathbb{1}_{3}$ is the identity matrix.
The NV spin's internal hyperfine coupling and quadrupole moment are neglected as we did not observed additional effects corresponding to these terms. 

The NV spin's ground state level structure is further modified by static electric ($\vec{E}$), magnetic ($\vec{B}$) and strain ($\vec{\delta}$) fields, whose contributions are given by
\begin{equation}
\begin{aligned}
\hat{\mathcal{V}}_{gs} = \: & {} \mu_{B} g^{\parallel }_{gs} \hat{S}_{z} B_{z}+
\mu_{B} g^{\bot }_{gs} \Big(\hat{S}_{x} B_{x}+\hat{S}_{y} B_{y}\Big)+
d^{\parallel }_{gs} (E_{z}+\delta_{z}) \Big[\hat{S}_{z}^{2}-S(S+1)/3~\mathbb{1}_{3}\Big]\\
& {} +d^{\bot }_{gs} (E_{x}+\delta_{x}) \Big(\hat{S}_{y}^{2}-\hat{S}_{x}^{2}\Big)+
d^{\bot }_{gs} (E_{y}+\delta_{y}) \Big(\hat{S}_{x} \hat{S}_{y}+\hat{S}_{y} \hat{S}_{x}\Big) ,
\end{aligned}
\end{equation}
where $\mu_{B}$ is the Bohr magneton, $g^{\parallel }_{gs}$ and $g^{\bot }_{gs}$ are the components of the ground state electronic g-factor tensor, $d^{\parallel }_{gs}$ and $d^{\bot }_{gs}$ are the components of the ground state electric dipole moment.
The electric field and reduced stress tensor terms are treated as a single effective electric field.  

%% EXCITED STATE HAMILTONIAN
The excited state of the nitrogen vacancy is an orbital doublet, which due to orbital averaging becomes a simplified three level system at higher temperatures.  
The low temperature fine structure of the NV spin is given by the effective Hamiltonian,
\begin{equation}
\begin{aligned}
\hat{\mathcal{H}}_{es} = \: & {} \mathbb{1}_{2} \otimes D^{\parallel }_{es}\Big[\hat{S}_{z}^{2}-S(S+1)/3~\mathbb{1}_{3}\Big]\\
& {} -\lambda^{\parallel }_{es} \: \hat{\sigma}_{y} \otimes \hat{S}_{z}
+D^{\bot }_{es}\Big[ \hat{\sigma}_{z} \otimes \Big(\hat{S}_{y}^{2}-\hat{S}_{x}^{2}\Big) - \hat{\sigma}_{x} \otimes \Big(\hat{S}_{x} \hat{S}_{y}+\hat{S}_{y} \hat{S}_{x}\Big) \Big]\\
& {} +\lambda^{\bot }_{es}\Big[ \hat{\sigma}_{z} \otimes \Big(\hat{S}_{x} \hat{S}_{z}+\hat{S}_{z} \hat{S}_{x}\Big) - \hat{\sigma}_{x} \otimes \Big(\hat{S}_{y} \hat{S}_{z}+\hat{S}_{z} \hat{S}_{y}\Big) \Big] ,
\end{aligned}
\end{equation}
where $\sigma_{x,y,z}$ are the standard two level Pauli spin matrices, $D_{es}^{\parallel }$ and $D_{es}^{\bot }$ are the spin-spin interaction terms, and  $\lambda^{\parallel }_{es}$ and $\lambda^{\bot }_{es}$ are mixing terms that arise from spin-orbit and spin-spin interactions. % Jero Email
This excited state Hamiltonian results in two spin-1 systems, one for each of the two orbital branches.

The influence of external fields on the NV spin's excited states level structure is given by,
\begin{equation}
\begin{aligned}
\hat{\mathcal{V}}_{es}^{LT} = \: & {} d^{\parallel }_{es}\Big(E_{z}+\delta_{z}\Big)~\mathbb{1}_{2} \otimes \mathbb{1}_{3}
+d^{\bot }_{es}\Big(E_{x}+\delta_{x}\Big)~\hat{\sigma}_{z} \otimes \mathbb{1}_{3}
-d^{\bot }_{es}\Big(E_{y}+\delta_{y}\Big)~\hat{\sigma}_{x} \otimes \mathbb{1}_{3}\\
& {} +\mu_{B} \: l^{\parallel }_{es} \: B_{z} \: \hat{\sigma}_{y} \otimes \mathbb{1}_{3}
+\mathbb{1}_{2} \otimes \mu_{B} \: g^{\parallel }_{es} \: B_{z}~\hat{S}_{z}
+\mu_{B}\: g^{\bot }_{es} \Big( B_{x} \hat{S}_{x} +  B_{y} \hat{S}_{y}\Big),
\end{aligned}
\end{equation}
where $d^{\parallel }_{es}$ and $d^{\bot }_{es}$ are components of the electronic dipole moment, $l^{\parallel }_{es}$ is the orbital magnetic moment also referred as $g_l$, the effective orbital g-factor and $g^{\parallel }_{es}$ and $g^{\bot }_{es}$ are components of the electronic g-factor tensor.

\subsection{Definition of the states and combined Hamiltonian}\label{Subsection:States}

In the simulations of the NV spin's energy levels we use a set of eigenstates that form an eigenbasis of the combined GS and ES \NVm{} Hamiltonians.
For simplicity we combine the spin singlet states into one state which then results in 10 states: 3 ground states, 6 excited states and one singlet state. 
The states are defined as:\\
Ground states:
\begin{equation*}	
\begin{aligned}
	\ket{1} &\equiv {^3}A_2^{-1} \\
	\ket{2} &\equiv {^3}A_2^0 \\
	\ket{3} &\equiv {^3}A_2^{+1}
\end{aligned}
\end{equation*}
Excited states:
\begin{equation*}	
\begin{aligned}[c]
	\ket{4} &\equiv {^3}E_y^{-1} \\
	\ket{5} &\equiv {^3}E_y^0 \\ 
	\ket{6} &\equiv {^3}E_y^{+1}
\end{aligned}
\qquad\qquad\qquad
\begin{aligned}[c]
	\ket{7} &\equiv {^3}E_x^{-1} \\
	\ket{8} &\equiv {^3}E_x^{0} \\
	\ket{9} &\equiv {^3}E_x^{+1}
\end{aligned}
\end{equation*}
Combined singlet state:
\begin{equation*}	
\begin{aligned}
	\ket{10} &\equiv {^1}A_1/{^1}E \\
\end{aligned}
\end{equation*}
with the notation of $X^{m_s}$, where the right superscript $m_s$ indicates the spin level. For the states ${^3}E$ the right subscript indicates the orbital branch, e.g. $E_y^{-1} \equiv \ket{E_y, -1} = \ket{\text{orbital state}, \text{spin state}}$. 
Following this definition, the combined Hamiltonian is a $10\times10$ matrix containing the Hamiltonians of both the ground and excited states.

% combined H
\newlength{\mycolwd}
\settowidth{\mycolwd}{$\hat{\mathcal{H}}_{gs}+\hat{\mathcal{V}}_{gs}$}
\newcommand\w[1]{\makebox[\mycolwd]{$#1$}}
\begin{equation}
\label{equation:CombinedH}
H = 2\pi
\begin{pmatrix}
   % First row
  \w{\hat{\mathcal{H}}_{gs}+\hat{\mathcal{V}}_{gs}}  & 
  \begin{matrix} 0 & & & \dots & \\ \vdots &  &  &  & \\ 0 & & & \dots & \end{matrix} &
  \begin{matrix} 0 \\ \vdots \\ 0 \end{matrix}\\
   % Second row
   \begin{matrix} 0 & & \dots & & 0\\ \\ \vdots & & & & \vdots \\  \end{matrix} &
   \w{\hat{\mathcal{H}}_{es}+\hat{\mathcal{V}}_{es}} &
   \begin{matrix} \vdots \end{matrix}\\
   % Third row
   \begin{matrix} 0 & & \dots & & 0 \end{matrix} &
   \begin{matrix} \dots \end{matrix} &
   \w{\begin{matrix} 0 \end{matrix}}
\end{pmatrix}
\end{equation}

\subsection{Optical transition rates}\label{Subsection:OpticalTransitionRates}

The NV$^{-}$ spin state can be optically excited from the ground state into excited state using green laser illumination, via a spin conserving dipole-allowed transition. 
Once in the excited state manifold, the NV spin can decay back to the ground state via either a radiative decay path ($k_r$) or a non-radiative ($k_{nr}$) one.
The radiative decay is a direct spin-conserving decay from the excited state back to the ground state. 
While the non-radiative decay path goes through the inter-system crossing (ISC) into a metastable singlet state, which exhibits different transition rates depending on the electron spin state, $m_s =0$ versus $m_s = \pm 1$ spin states\,\cite{Tetienne2012}.

In the following we make the assumption that all spin conserving transition rates from the excited to ground states are the same independent of electron spin state ($k_{\ket{4,5,6} \to \ket{1,2,3}} = k_r$) and that the non-spin conserving transitions are zero  (e.g. $k_{\ket{6} \to \ket{1}} = 0$).
The non-spin conserving transitions rates have previously been shown to only be a few percent compared to the spin-conserving one\,\cite{Robledo2011a}.

The transition rates from the ground to excited states and the corresponding decay transition rates from the excited to ground states are thus defined as
\begin{equation}	
\begin{aligned}[c]
k_{\ket{1} \to \ket{4}} &=\beta_{E_{y}} \: k_{r}\\
k_{\ket{2} \to \ket{5}} &=\beta_{E_{y}} \: k_{r}\\
k_{\ket{3} \to \ket{6}} &=\beta_{E_{y}} \: k_{r}
\end{aligned}
\qquad\qquad
\begin{aligned}[c]
k_{\ket{1} \to \ket{7}} &=\beta_{E_{x}} \: k_{r}\\
k_{\ket{2} \to \ket{8}} &=\beta_{E_{x}} \: k_{r}\\
k_{\ket{3} \to \ket{9}} &=\beta_{E_{x}} \: k_{r}
\end{aligned}
\qquad\qquad
\begin{aligned}[c]
k_{\ket{4} \to \ket{1}} &=k_{r}\\
k_{\ket{5} \to \ket{2}} &=k_{r}\\
k_{\ket{6} \to \ket{3}} &=k_{r}
\end{aligned}
\qquad\qquad
\begin{aligned}[c]
k_{\ket{7} \to \ket{1}} &=k_{r}\\
k_{\ket{8} \to \ket{2}} &=k_{r}\\
k_{\ket{9} \to \ket{3}} &=k_{r}
\end{aligned}
\end{equation}

where $\beta_{E_{x}}$ and $\beta_{E_{y}}$ are the pumping parameters which are proportional to laser power and capture the polarization dependence of the excitation, and $k_{r}$ is the respective relaxation rates from the excited state to the ground state.

% Decay into singlet state and decay from singlet into the GS
The transition rates from the excited states to the metastable state are spin dependent whereas the rates from the metastable state to the ground states are similar for all spin states. These rates are defined as
\begin{equation}
\label{equation:DecayRates}
\begin{aligned}[c]
k_{\ket{4} \to \ket{10}} &=k_{nr_{\pm 1}}\\
k_{\ket{5} \to \ket{10}} &=k_{nr_{0}}\\
k_{\ket{6} \to \ket{10}} &=k_{nr_{\pm 1}}
\end{aligned}
\qquad\qquad\qquad
\begin{aligned}[c]
k_{\ket{7} \to \ket{10}} &=k_{nr_{\pm 1}}\\
k_{\ket{8} \to \ket{10}} &=k_{nr_{0}}\\
k_{\ket{9} \to \ket{10}} &=k_{nr_{\pm 1}}
\end{aligned}
\qquad\qquad\qquad
\begin{aligned}[c]
k_{\ket{10} \to \ket{1}} &=k_{m_{\pm 1}}\\
k_{\ket{10} \to \ket{2}} &=k_{m_{0}}\\
k_{\ket{10} \to \ket{3}} &=k_{m_{\pm 1}}
\end{aligned}
\end{equation}
where the optical spin contrast of the NV results from $ k_{nr_{0}} \ll k_{nr_{\pm 1}}$\,\cite{Tetienne2012} and  the decay rates from the metastable rates are approximately equal $k_{m_{0}} \approx k_{m_{\pm 1}}$.

%% Effect of MW field
In a typical optically detected magnetic resonance (ODMR) measurement the spin population is transferred between the spin states of the ground state via an applied microwave field, and the corresponding transition rates are defined as
\begin{equation}
\begin{aligned}[c]
k_{\ket{1} \to \ket{2}} &=k_{\ket{2} \to \ket{1}}=k_{MW_{-1}}
\end{aligned}
\qquad\qquad\qquad
\begin{aligned}[c]
k_{\ket{3} \to \ket{2}} &=k_{\ket{2} \to \ket{3}}=k_{MW_{1}}
\end{aligned}
\end{equation}
where $k_{MW_x}$ is the driving transition rate on resonance with the $x$ transition between the $\ket{2}$ and  $\ket{x}$ states.
The transition rates $k_{\ket{1} \to \ket{2}}$ and $k_{\ket{3} \to \ket{2}}$ are in general zero in our experiments because no microwave (MW) driving field is applied.
These rates are only set to non-zero when one wants to model the effect of the level anti crossings on the spin-readout contrast, which is modelled in Section~\ref{Subsection: ODMR}. 

\begin{table}
	\begin{tabular}{l  l  l  l  l  l}
		\hline
		Reference &  $k_{r}$ \quad\quad & $k_{nr_{0}}$ \quad\quad & $k_{nr_{1}}$ \quad\quad & $k_{m_{0}}$ \quad\quad &  $k_{m_{\pm 1}}$ \quad\\
		\hline 
		 Robledo et al.\,\cite{Robledo2011b} \quad\quad & 65 & 11 & 80 & 3.0 & 2.6 \\
		 Tetienne et al.\,\cite{Tetienne2012} \quad\quad & 65.9 & 7.9 & 53.3 & 0.98 & 0.73 \\
		 Gupta et al.\,\cite{Gupta2016a} \quad\quad & 66.8 & 10.5 &90.7 &4.8 &2.2\\
		 \hline
	\end{tabular}
\caption{Experimentally measured transition rates at zero field. All of the rates are in MHz.}
\label{table:rates}
\end{table}

All the transition rates in Eq.~\ref{equation:DecayRates} have been measured experimentally, shown in Table\,\ref{table:rates}. 
In the model we used the parameters from Gupta et al.\,\cite{Gupta2016a}, which we found to be in best agreement with our data.

\subsection{Phonon-induced transition rates between the excited state orbitals}\label{Subsection:OrbitalTransitionRates}

The interaction of the electron orbital in the excited state with $A_1$ and $E$ symmetry phonons, up to the first order, can be written as~\cite{Fu2009a, Plakhotnik2015a}
\begin{equation}
H_{ep}=\sum_{i}{\hbar \lambda^A_i V^A (a^A_i+a^{A\dagger}_i)}+\sum_{i}{\hbar \lambda^E_i \left[V^E_x (a^E_{i,x}+a^{E\dagger}_{i,x})-V^E_y (a^E_{i,y}+a^{E\dagger}_{i,y})\right]},
\end{equation}
where
\begin{eqnarray}
V^A=|X\rangle\langle X|+|Y\rangle\langle Y|,\\
V^E_x=|X\rangle\langle X|-|Y\rangle\langle Y|,\\
V^E_y=|X\rangle\langle Y|+|Y\rangle\langle X|,
\end{eqnarray}
are the electron orbital operators for interaction with $A_1$, $E_x$ and $E_y$ phonon modes, respectively.
The orbitals are denoted X and Y with $\ket{X}=\big(\begin{smallmatrix}1\\0\end{smallmatrix}\big)$ and $\ket{Y}=\big(\begin{smallmatrix}0\\1\end{smallmatrix}\big)$, $\lambda^A_i$ and $\lambda^E_i$ are the electron-phonon coupling coefficients, and $a^M_{i,p}$ ($a^{M\dagger}_{i,p}$) are the annihilation (creation) operators of the $i$-th $M=A_1, E$ phonon modes.

Using the Fermi golden rule (FGR) we can calculate the phonon induced transition rates between the excited states.
The transition rate between the states $|i\rangle$ and $|f\rangle$ with energies $E_i$ and $E_f$, respectively, is given by 
\begin{equation}
    \Gamma_{if}=\frac{2\pi}{\hbar}\left|{(H^{(2)}_{ep})_{fi}+\sum_{m}{\frac{(H_{ep})_{fm} (H_{ep})_{mi}}{E_i-E_m}}}\right|^2\delta(E_i-E_f),
\end{equation}
where $H^{(2)}_{ep}$ is the second order electron-phonon Hamiltonian, which we expect to be negligible. The summation in the second term is over all intermediate orbital states. 

Because only resonant phonons can drive spin transitions directly via one-phonon absorption or emission,
we expect that the dominant mechanism is the two-phonon Raman processes of the form of absorption of a phonon followed by emission of a phonon or emission followed by absorption.
Considering the initial and final phonon states $|n_k,n_l\rangle$, and $|n_k+1,n_l-1\rangle$, respectively, the transition rate from $|X\rangle$ to $|Y\rangle$ is given by
\begin{equation}
\begin{aligned}
\Gamma_{XY} = &2\pi\hbar^3\sum_{k,l} \left|\frac{\lambda^E_l (\lambda^A_k +\lambda^E_k)}{(-\hbar \omega_k)}+\frac{\lambda^E_k (\lambda^A_l +\lambda^E_l)}{\hbar \omega_l}+\frac{\lambda^E_k (\lambda^A_l -\lambda^E_l)}{E_x - E_y -\hbar \omega_k}+\frac{\lambda^E_l (\lambda^A_k -\lambda^E_k)}{E_x - E_y +\hbar \omega_l}\right|^2\\
& \times n_l (n_k +1) \delta(E_x - E_y + \hbar \omega_l - \hbar \omega_k). 
\end{aligned}
\end{equation}
Converting the summations to integrals and using the phonon spectral density for M=A$_1$ or M=E phonon modes
\begin{equation}
J_M (\hbar \omega)=\sum_{k}(\hbar \lambda^{M}_k)^2 \delta(\hbar \omega -\hbar \omega_k),
\end{equation}
we can then write
\begin{equation}
\Gamma_{XY}=\Gamma^{EE}_{XY}+\Gamma^{AE}_{XY},
\end{equation}
where 
\begin{equation}
\begin{aligned}
\Gamma^{M E}_{XY} = \frac{4\pi}{\hbar}&\int{\left[\frac{J_M (\Delta_{xy}+\hbar \omega) J_E (\hbar \omega)}{(\Delta_{xy}+\hbar \omega)^2}+\frac{J_E (\Delta_{xy}+\hbar \omega) J_M (\hbar \omega)}{(\hbar\omega)^2} \right]} \\
& \times n(\hbar \omega)[n(\hbar \omega+\Delta_{xy})+1]~d(\hbar \omega). % Still in integral but linebreak requires closed inegral
\end{aligned}
\end{equation}
Here, $M=A_1, E$ are the phonon symmetries, $\Delta_{xy}=E_x - E_y=2\delta_{\perp}=2\sqrt{\delta_{x}^{2}+\delta_{y}^{2}}$, and $n(\hbar \omega)=(e^{\hbar\omega/(k_B T)}-1)^{-1}$ is the mean number of phonons at thermal equilibrium.
For this calculation we have used random phase approximation \cite{Cambria2022}.

For long wavelength acoustic phonons, we have \cite{Goldman2015b}
\begin{equation}
J_{M} (\hbar \omega)=\eta_M (\hbar \omega)^3,
\end{equation}
where $\eta_M$ is the coupling strength between the electron states and M-symmetric acoustic phonons.

Combining the equations from above results in an analytical expression of the transition rate $\Gamma_{XY}$, which will be used in the model. It reads
\begin{equation}\label{Eq:phonon_coupling_rate}
\Gamma_{XY}=\frac{4\pi}{\hbar}(\eta^2_E+\eta_A \eta_E) (k_B T)^5 I(\omega_{\textrm{c}},\delta),
\end{equation}
where 
\begin{equation}
\label{equation:Integral}
I(\omega_{\textrm{c}},T,\delta_{\perp})=\int^{x_{\textrm{c}}}_{0}dx~x (x+x_{\Delta})\left[x^{2}+(x+x_{\Delta})^{2}\right]\frac{1}{(e^{x} -1)(1-e^{-(x+x_{\Delta})})},
\end{equation}
with $x_{\textrm{c}}=\hbar\omega_{\textrm{c}}/(k_B T)$, $\omega_{\textrm{c}}$ being the cutoff frequency in the integral, $x=\hbar \omega/(k_B T)$, and $x_{\Delta}=\Delta_{xy}/(k_B T)=2\sqrt{\delta^2_x+\delta^2_y}/(k_B T)$.

In contrast to other work, $\omega_{\textrm{c}}$ is not the Debye frequency.
The Debye model is an over simplification of all modes in a crystal.
It considers only an acoustic type of phonons with a cutoff frequency which is obtained in such a way that the area below the density of states as a function of the mode frequency is the same as that of the real crystal.
Therefore, the Debye model erases all particularities of the real modes:
the positions (frequencies) at which the density of states achieves its maxima, the maximum phonon frequency, optical phonons, etc.
The best simplest approximation is one that considers (1) an acoustic branch with a cutoff frequency that matches the frequency at which the density of states achieves a maximum for the acoustic branch (60 meV in our case); 
and (2) an optical branch with cutoff frequencies (for optical phonons there is a low and high frequency cutoff) that matches the frequencies at which the density of states achieves maxima slightly above the maximum of the acoustic branch that is mentioned before and the maximum of the optical branch (which turns to the maximum optical phonon frequency of the real crystal).
However, in this model, we chosen just the acoustic part for two reasons: (A) to keep it simple and because as temperature rises, the acoustic phonons (from low frequencies up to the frequency at which the density of states (DOS) achieves its acoustic maximum (60 meV) start to matter more than the optical phonons.

In the model we combine some parts of Eq.~\ref{Eq:phonon_coupling_rate} into an effective coupling $\alpha_{\textrm{ph}}$ and set $\omega_{\textrm{c}}$ to 60~meV
which results to
\begin{equation}
\label{Eq:phonon_coupling_prefactor}
\Gamma_{XY} = \alpha_{\textrm{ph}} I(\omega_{\textrm{c}}=\textrm{60~meV},T,\delta_{\perp}) T^5.
\end{equation}
The transition rate from orbital $X$ to orbital $Y$ and vice versa is then defined as
\begin{equation}
k_{\ket{Y} \to \ket{X}} = k_{\ket{X} \to \ket{Y}} = \alpha_{\textrm{ph}} I(\omega_{\textrm{c}}=\textrm{60~meV},T,\delta_{\perp}) T^5.
\end{equation}

\subsection{Lindblad master equation}\label{Subsection:Lindblad}

We employ the Lindblad master equation to model the time-evolution of the \NVm{} Hamiltonian under the influence of relaxation, laser-induced optical pumping and phonon-induced orbital averaging.
For an introduction and a derivation refer to Ref.~\cite{Manzano2020}.

In general, the Lindblad master equation is a simplified method for describing the evolution of different types of open quantum systems which are weakly coupled to an environment. 
It tries to capture the influence of this weakly coupled environment on the system in a simplified form, the Lindblad operator ($L$). 
This avoids solving the full Hamiltonian that contains all the additional quantum and semi-classical interactions of the target quantum system and the bath that it is connected to, which in many cases is not even possible. 
In this fashion, it reduces this complexity to a series of imposed decay rates, which is commonly used to model dephasing and relaxation of quantum states.

Under the assumptions of Markovianity and time-homogeneity, a special type of the Lindblad master equation, the Markovian master equation, describes the evolution of the density matrix ($\rho$) of our combined system\cite{Preskill1998}:

\begin{equation}
\label{eq:masterequation}
\frac{d\rho}{dt}=-\frac{i}{\hbar}[H,\rho]+\sum^{}_{l}{\mathcal{D}(L_l)\rho},
\end{equation}
where $H$ the Hamiltonian of the system and
\begin{equation}
\mathcal{D}(L_{l})\rho=L_{l}\rho L^{\dagger}_{l}-\frac{1}{2}\left(\rho L^{\dagger}_{l} L_{l} + L^{\dagger}_{l} L_{l} \rho\right),
\end{equation}
where $L_l$ the Lindblad operator that represents the $l^{\rm th}$ interaction with the bath. 

The first term in the master equation describes the unitary evolution of the density matrix due to the Hamiltonian of the \NVm{} system, while the second term describes the evolution due to the interaction with the environment. 
The Lindblad operators $L_l$ result in transitions between states of the system due to the interaction with the environment. 
Each operator, $L_l=\sqrt{\Gamma_{l}}A_{l}$, describes a different aspect of the environment where $\Gamma_{l}$ describes the strength of the interaction and can be a relaxation rate, dephasing rate, etc.
When all $L_{l}$ are zero, the Liovillian equation of a closed quantum system is recovered. 

\subsubsection{Lindblad operators}

In a next step we define the Lindblad operators $L_{l}$ which describe the effects we include in our model.
First we look at how to incorporate the relaxation and laser-induced optical pumping rates from Section~\ref{Subsection:OpticalTransitionRates} and then continue to the phonon-induced transitions introduced in Section~\ref{Subsection:OrbitalTransitionRates}.

In general the Lindblad operator representing the transition from state $\ket{i}$ to $\ket{j}$ is given by
\begin{equation}
L_{\ket{i} \to \ket{j}}=\sqrt{k_{\ket{i} \to \ket{j}}}\ket{j}\bra{i},
\end{equation}
where $k_{\ket{i} \to \ket{j}}$ is the transition rate from state $\ket{i}$ to $\ket{j}$.
This applies for all transition rates in Section ~\ref{Subsection:OpticalTransitionRates}. 

\textbf{Phonon induced transitions}

The Lindblad operators from phonon induced transitions for the different excited state orbital doublet ($X$ and $Y$) are constructed slightly different.
We are concerned with transitions within the excited state manifold, so the states in question are the subset of $\ket{i,j}=\{4\dots9\}$
The two orbitals are defined as $\ket{X}=\big(\begin{smallmatrix}1\\0\end{smallmatrix}\big)$ and $\ket{Y}=\big(\begin{smallmatrix}0\\1\end{smallmatrix}\big)$. 
We define the spin conserving transition operator from the upper orbital to the lower orbital as
\begin{equation}
L_{\downarrow}=\sqrt{k_{\ket{Y} \to \ket{X}}}\ket{Y}\bra{X} \otimes \mathbb{1}_{3},
\end{equation}
where $k_{\ket{Y} \to \ket{X}}$ is the transition operator from the upper orbital $Y$ to the lower orbital $X$. 
The operator for the spin conserving rate from the lower to the upper orbital $L_{\uparrow}$ is equivalent with flipped $X$ and $Y$ orbitals and indices.

\subsubsection{Vectorization for application of Super Operators}

We now introduce a vectorization procedure that maps $\ket{i} \bra{j} \mapsto \ket{j} \otimes \ket{i}$, such that it is possible to use the Super Operator formalism.\,\cite{Scopa2019}
The vector form of the density matrix is obtained by stacking the columns of an $n\times n$ density matrix from left to right on top of each other to form a vector of length $n^2$, such that
\begin{equation}
\hat\rho = \sum_{i,j}{\rho_{i,j}\ket{j} \otimes \ket{i}}.
\end{equation}
Using this vectorization, we can rewrite the Lindblad master equation from Eq.~\ref{eq:masterequation} as a product between a matrix and a vector, where all the properties are contained in the Lindbladian superoperator, given by
\begin{equation}
\mathcal{\hat L}=i(\bar H\otimes \mathbb{1}_{10}-\mathbb{1}_{10} \otimes H)+\sum^{}_{l}{\left(\bar L_{l}\otimes L_{l}-\frac{1}{2}\mathbb{1}_{10}\otimes L^{\dagger}_{l} L_{l}-\frac{1}{2} \bar L^{\dagger}_k \bar L_k\otimes \mathbb{1}_{10}\right)}.
\end{equation}
where the complex conjugate is shown with an over bar (e.g.: $\bar H$) and the adjoint with a dagger ($\dagger $). $l$ denotes all possible transitions, both optical and phonon-induced.

The time-evolution of the system in the vectorized form of the master equation~\cite{Havel2003} is given by
\begin{equation}
\frac{d \hat\rho(t)}{dt}= \mathcal{\hat L} \hat\rho(t),
\end{equation}
which is computationally more efficient to calculate.

\subsection{Time-dependent and steady state solution of the master equation}

In the following we will first provide a general time-dependent solution to the Lindblad master equation and then proceed to the steady state solution.
When comparing the two approaches they result in very similar results.

\textbf{Time evolution of the density matrix}

The time evolved density matrix is obtained by
\begin{equation}
\label{equation:TimeEvolutionLindblad}
\hat \rho (t)= e^{\mathcal{L}t}\hat\rho(0) \equiv e^{\mathcal{L}t}\hat\rho_{0},
\end{equation} 
where $\hat{\rho_{0}}$ is the vectorised initial density matrix of the system.
We choose an initial population in the three states of the ground state of the \NVm{}:
\begin{equation}
\rho_{0} = \frac{1}{3}\ket{1}\bra{1} + \frac{1}{3}\ket{2}\bra{2} + \frac{1}{3}\ket{3}\bra{3},
\end{equation}
which is then converted to its vectorized form $\hat\rho_{0}$ as discussed in the previous section.

We solve Eq.~\eqref{equation:TimeEvolutionLindblad} with a time of \SI{3}{\micro \second} which is enough under continuous laser excitation to obtain a steady state. We obtain the vectorized density matrix $\hat \rho (t)$ which we convert back to its matrix form. We then extract the populations of the 10 states by
\begin{equation}
p_{i} = \bra{i}\rho(t)\ket{i}.
\end{equation}

\textbf{Steady state solution of the density matrix}

In an open quantum system where the decay rates are larger than the corresponding excitation rates, there is a steady state solution for the density matrix for $t \to \infty$\,\cite{Nation2015}. 
In this case the solution of the Lindblad master equation can be cast into an eigenvalue equation:
\begin{equation}
\label{equation:SteadyStateEigenvalue}
\mathcal{\hat L}\hat\rho_{ss} = 0\hat\rho_{ss},
\end{equation}
with $\hat\rho_{ss}$, the vectorized form of the steady state density matrix, and $\mathcal{\hat L}$, the Lindbladian superoperator in the given basis.

We can solve equation \ref{equation:SteadyStateEigenvalue} and convert the vectorized density matrix back to its matrix form. 
The steady state populations of the 10 states then read after normalization
\begin{equation}
p_{i} = \frac{\bra{i}\rho_{ss}\ket{i}}{\sum_{j=1}^{10} \bra{j}\rho_{ss}\ket{j}}.
\end{equation}

\subsection{Photoluminescence}

The photoluminescence of the \NVm{} ($I_{\textrm{PL}}^-$) is calculated by summing over the relevant radiative transitions from the excited state to the ground state and the populations of these states, such that 
\begin{equation}
I_{\textrm{PL}}^-=\sum_{i=4\dots9}k_{r}p_{i}.
\end{equation}

Additional modification to the photoluminescence such as background fluorescence, $I_{\textrm{bck}}$, and collection efficiency, $\eta_{\textrm{collection}}$, can also be introduced, such that 
\begin{equation}
	I^{-}_{\textrm{Total}} = \eta_{\textrm{collection}} I_{\textrm{PL}}^- + I_{\textrm{bck}},
\end{equation}
which for convenience is referred to as $I_{\textrm{PL}}^-$ elsewhere in the main manuscript and supplementary information.

From the photoluminescence the ODMR contrast $\mathcal{C}$ can be calculated by including non-zero microwave transition rates into the Hamiltonian, i.e. $k_{MW} \neq 0$. For example, with a microwave driving between $\ket{2}$ (${^3}A_2^0$) and $\ket{3}$ (${^3}A_2^{-1}$), the contrast can be determined by calculating
\begin{equation}\label{Eq: contrast}
\begin{aligned}[c]
\mathcal{C} =\frac{I^{-}_{\textrm{PL}}(k_{MW_{-1}}=0) -I^{-}_{\textrm{PL}}(k_{MW_{-1}}\neq0)}{I_{\textrm{PL}}(k_{MW_{-1}}=0)}.
\end{aligned}
\end{equation}

\section{Experimental setup}
\label{Subsection:Setup}

The measurements were performed with a commercial confocal microscope in a closed-cycle cryostat (attoDRY2200, Attocube).
The cryostat has a variable temperature range of $T = 2 - 300$~K.
An integrated PID controller is used to stabilize the sample space temperature using two sets of heaters, one directly below the sample and another one heating the VTI (Variable Temperature Inset). 
The VTI has two different operating temperature regimes ($T < 15$~K and $T > 15$~K).
The temperature sensor at the sample is a thin film resistance temperature sensor (Cernox CX-1050-SD-HT-1z4L, Lakeshore). 
Over the complete range the temperature stability is generally better than $\Delta T \pm 100$~mK, but occasional drifts occur.

The NV is optically excited using a 532~nm laser (Torus, Laser Quantum) and a fibre-coupled AOM (Fibre-Q, EQ Photonics).
The photoluminescence of the NV is separated using a dichoric mirror (DMLP567R, Thorlabs), filtered by a 650nm long pass filter (FELH0650, Thorlabs) and recorded on an avalanche photodiode (SPCM-AQRH-33, Excelitas).

The NV spin state is controlled via microwaves that are produced by a signal generator (SMBV100B, Rhode Schwartz), controlled with a the signal generator's inbuilt IQ mixer, and subsequently amplified (HPA-25W-63+, Minicircuits).
The microwaves are applied to the NV using a lithographically pattered structure on the diamond.

The cryostat has a superconducting vector magnet (1~T,1~T,1~T) with full 1~T vectorial control, operated by two magnet controller (APS100, Attocube).

\section{Samples}
\label{sec:Samples}

In this work we investigate native and in-grown single nitrogen vacancy centers in bulk diamond using photonic structures.
We look at two different samples in this work, one containing NVs with an elevated effective strain field and one containing NVs with a low effective strain field.
On both samples we fabricated microscopic diamond Solid-Immersion-Lenses (SIL) to increase photon collection efficiency. 
The SILs were created with focussed ion beam (FIB) milling according to Ref.\,\cite{Jamali2014a} with a radius of \SI{4.2}{\micro\meter}.
The samples were acid cleaned using a well-established acid cleaning technique\cite{Brown2019} which leaves the diamond surface predominantly O-terminated.
Afterwards an antenna structure was pattered on top of the diamond next to the SILs.  

\textbf{Sample containing NVs with elevated effective strain: }
The sample is identical to the sample used in Ref.\,\cite{Neu2014a,Happacher2022a}, where extensive details on the sample and sample fabrication are given.
The sample was CVD grown by the group of J. Achard and A. Tallaire with methods described in detail in Ref.\,\cite{Tallaire2014}. 
NV centers were created during the diamond growth through controlled incorporation of N gas into the growth reactor. 
The growth conditions used here lead to NV centers whose quantization axis is preferentially aligned with the $(100)$ growth direction and which typically show excellent spin properties\,\cite{Lesik2014a}. 
We note that the sample exhibits elevated fluorescence background levels from SiV centers which were inadvertently introduced during sample growth.

\textbf{Sample containing NVs with very low effective strain: } The sample is a thin $(100)-$oriented, chemical vapor deposition (CVD) grown type IIa ``electronic grade'' diamond from Element 6.
This diamond contains a slightly larger native N concentration than usual throughout the diamond.
We investigate naturally occurring NV centers.

\section{Experimental data}

\begin{figure}
	\centering
	\includegraphics[width=1\linewidth]{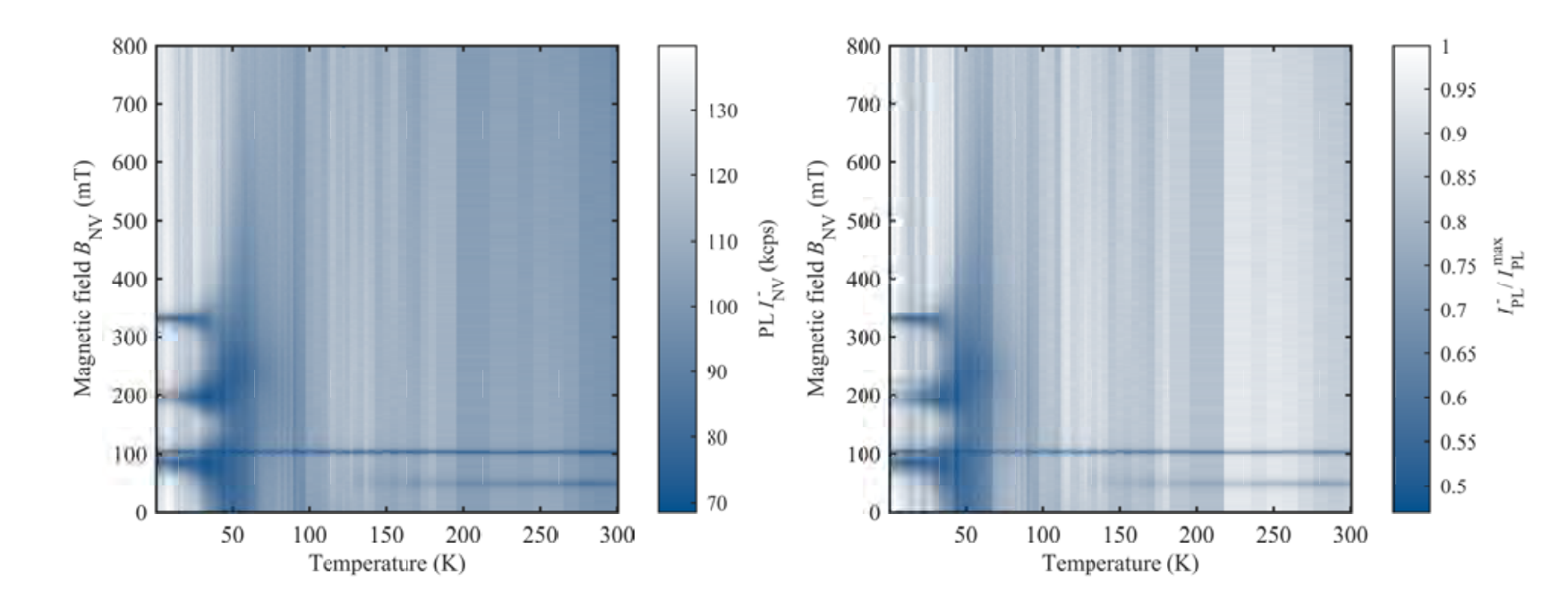}
	\caption{
    PL intensity $I_{\rm PL}^{-}$ as a function of $B_{\rm NV}$ and $T$ for NV \#1 for magnetic fields up to 800~mT. On the left the raw data of the measurement is shown. During the measurement the green excitation laser experienced strong drifts which lead to a distortion of the data. On the right the data is corrected with the simultaneously recorded laser power.
	}
	\label{fig:appendix_DataLowStrainNV}
\end{figure}

\begin{figure}
	\centering
	\includegraphics[width=1\linewidth]{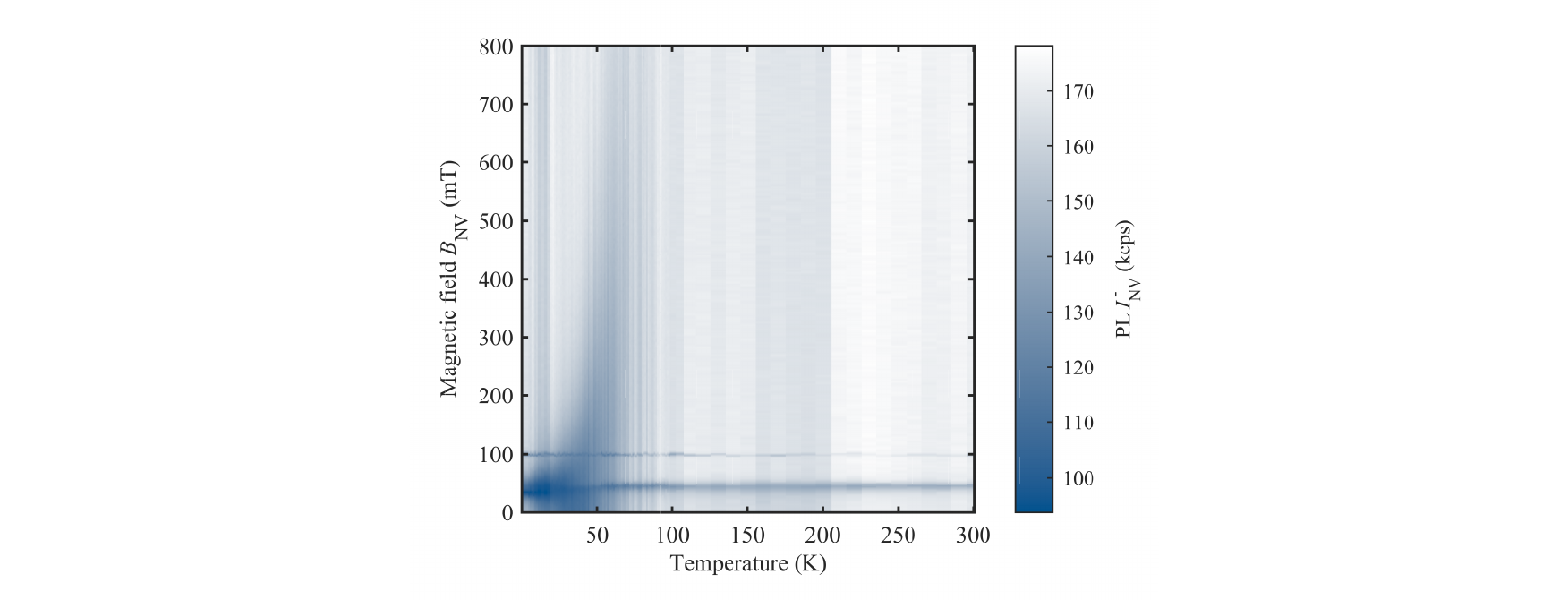}
	\caption{
    PL intensity $I_{\rm PL}^{-}$ as a function of $B_{\rm NV}$ and $T$ for NV \#2 with elevated effective strain for magnetic fields up to 800~mT.
	}
	\label{fig:appendix_DataHighStrainNV}
\end{figure}

The data in this work is collected by measuring the photoluminescence of two single NV in the two different samples described in section~\ref{sec:Samples}.
The measurements are taken through performing a magnetic field sweep at every recorded temperature. 
The results without any normalization are shown in Fig.~\ref{fig:appendix_DataLowStrainNV} and ~\ref{fig:appendix_DataHighStrainNV}. 

\subsection{Additional temperature sweeps}

\begin{figure}
	\centering
	\includegraphics[width=1\linewidth]{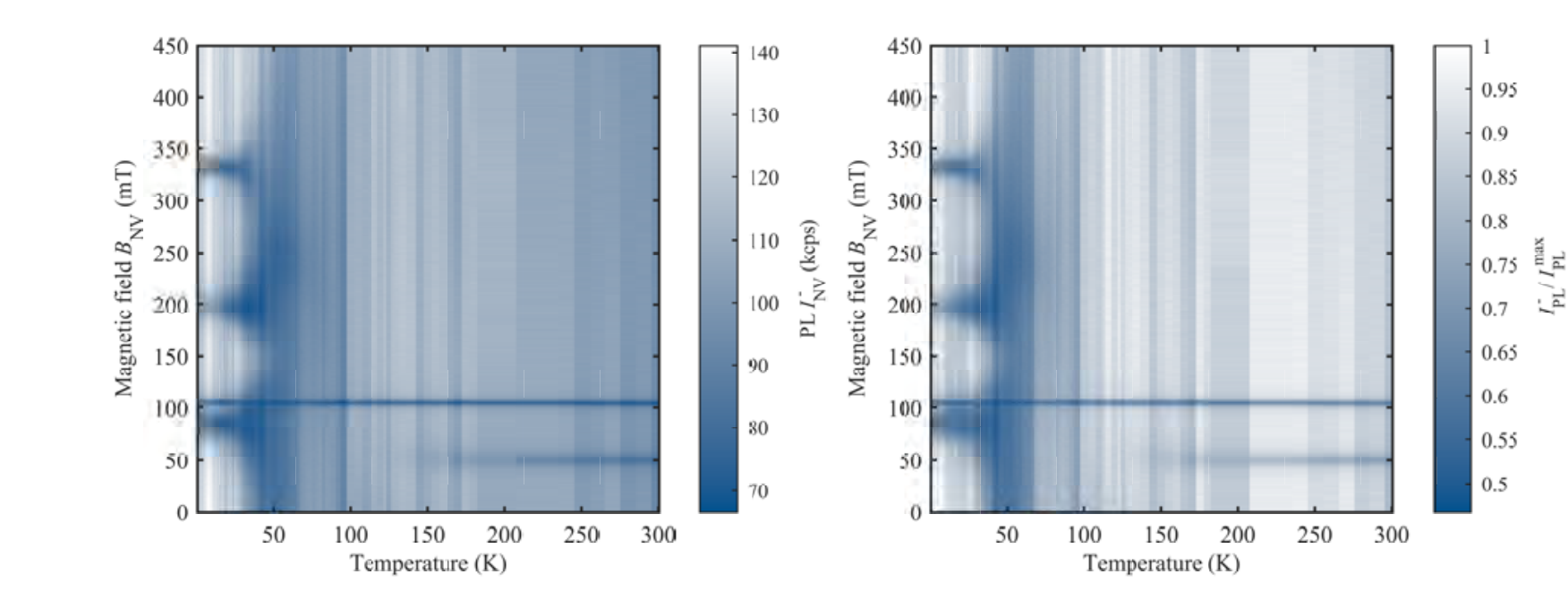}
	\caption{
		Additional PL intensity $I_{\rm PL}^{-}$ as a function of $B_{\rm NV}$ and $T$ measurement of NV \#1 with a smaller magnetic field range up to 450~mT and longer integration times. The raw data is shown on the left and the data corrected for laser drifts on the right.
	}
	\label{fig:appendix_DataLowStrainNV_2}
\end{figure}

\begin{figure}
	\centering
	\includegraphics[width=1\linewidth]{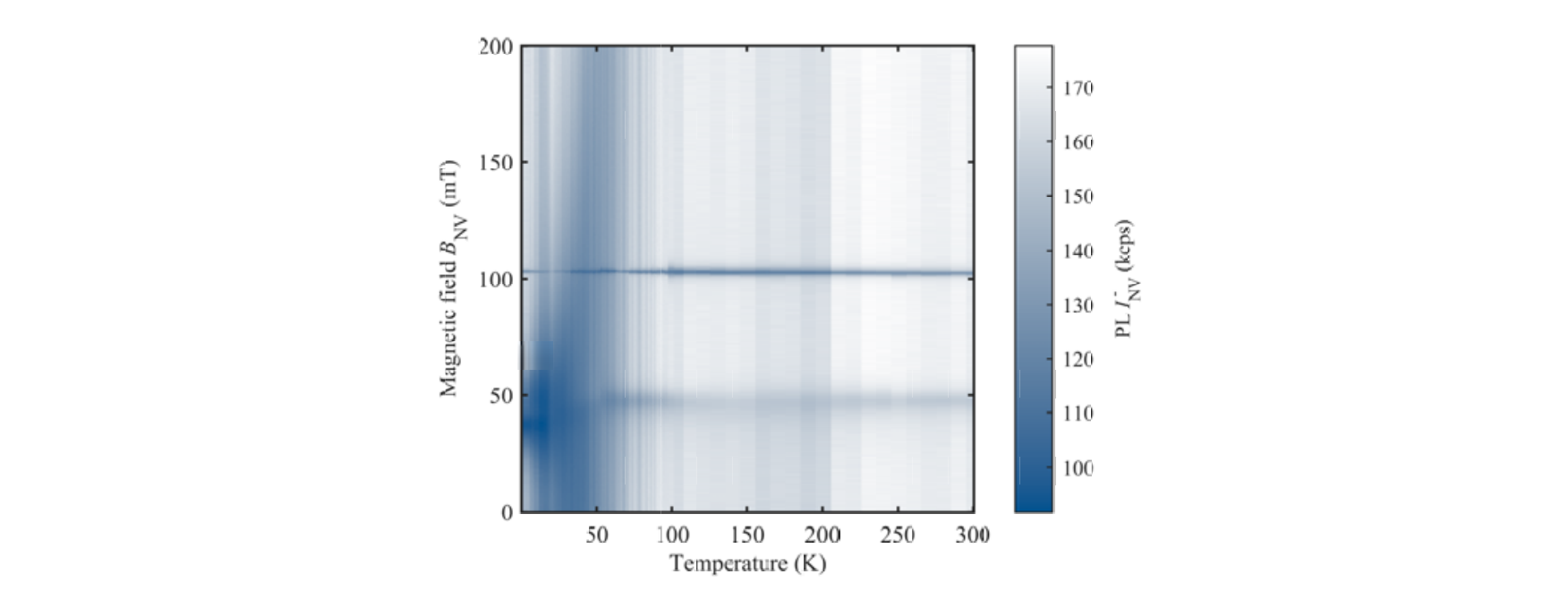}
	\caption{
		Additional PL intensity $I_{\rm PL}^{-}$ as a function of $B_{\rm NV}$ and $T$ measurement of NV \#2 for a smaller magnetic field range up to 200~mT and longer integration times.
	}
	\label{fig:appendix_DataHighStrainNV_2}
\end{figure}

Two additional datasets with longer integration times and a smaller magnetic field range are shown in Fig.~\ref{fig:appendix_DataLowStrainNV_2} and ~\ref{fig:appendix_DataHighStrainNV_2}.
These additional datasets have been acquired during the same experimental run as the other datasets.
At each temperature the magnetic field sweeps for the respective dataset have been measured consecutively.
Therefore they should only minimally differ in magnetic field misalignment and in illumination condition for each specific temperature.
The time for each magnetic field sweep varies between the datasets.
It is between $t = 300$~s and $t = 800$~s.

\subsection{Data normalisation}
\label{Subsection:DataNormalization}

During the measurement series we experienced unavoidable laser fluctuations which introduce significant noise to the measurement. 
This issue was particularly prevalent for the low strain measurement series. 
For one of the NVs (NV \#1) the laser power was monitored during the measurement series and therefore we were able to remove these additional artefacts.
The laser power was used as a normalisation factor that is independent of changes to the NV photo-physics as a function of temperature.
This normalisation process is depicted in Fig.~\ref{fig:appendix_DataLowStrainNV} and Fig.~\ref{fig:appendix_DataLowStrainNV_2}, where the left panel is the raw data and the right is after correcting for laser fluctuations.

Another way to normalize the data  is using the model itself as outlined below.
To normalize the data we use the extracted value for the effective coupling $\alpha_{\textrm{ph}}$ from Section~\ref{Subsection:FittingAlpha} and the intrinsic values of NV \#1 and NV \#2 to simulate the PL vs temperature $T$ at $B=800$~mT for each of the two NVs:
\begin{equation}
I_{\textrm{PL}\,\,\textrm{simulation}}^-(B=800~\textrm{mT},T)
\end{equation}
We choose this magnetic field value, because at higher magnetic field values the PL intensity is less dependent on the specific NV properties.
The normalization factor for the dataset of each individual NV is
\begin{equation}
\nu(T)=\frac{I_{\textrm{PL}\,\,\textrm{simulation}}^-(B=800~\textrm{mT},T)}{I_{\textrm{PL}\,\,\textrm{data}}^-(B=800~\textrm{mT},T)},
\end{equation}
where the normalized data is
\begin{equation}
I_{\textrm{PL}\,\,\textrm{normalized data}}(T) = \nu(T) I_{\textrm{PL}\,\,\textrm{data}}.
\end{equation}

Additionally, the temperature series was not equally spaced, as most of the features are present at lower temperatures. 
Leading to a temperature resolution of mostly $\Delta T = 1$~K at temperatures below $T < 100 K$ and a resolution of $\Delta T = 5-10$~K at higher temperatures.
For plotting the data the pixel sizes are changed to account for this change in resolution.

\subsection{Determining the intrinsic NV properties}

\begin{figure}
	\centering
	\includegraphics[width=1\linewidth]{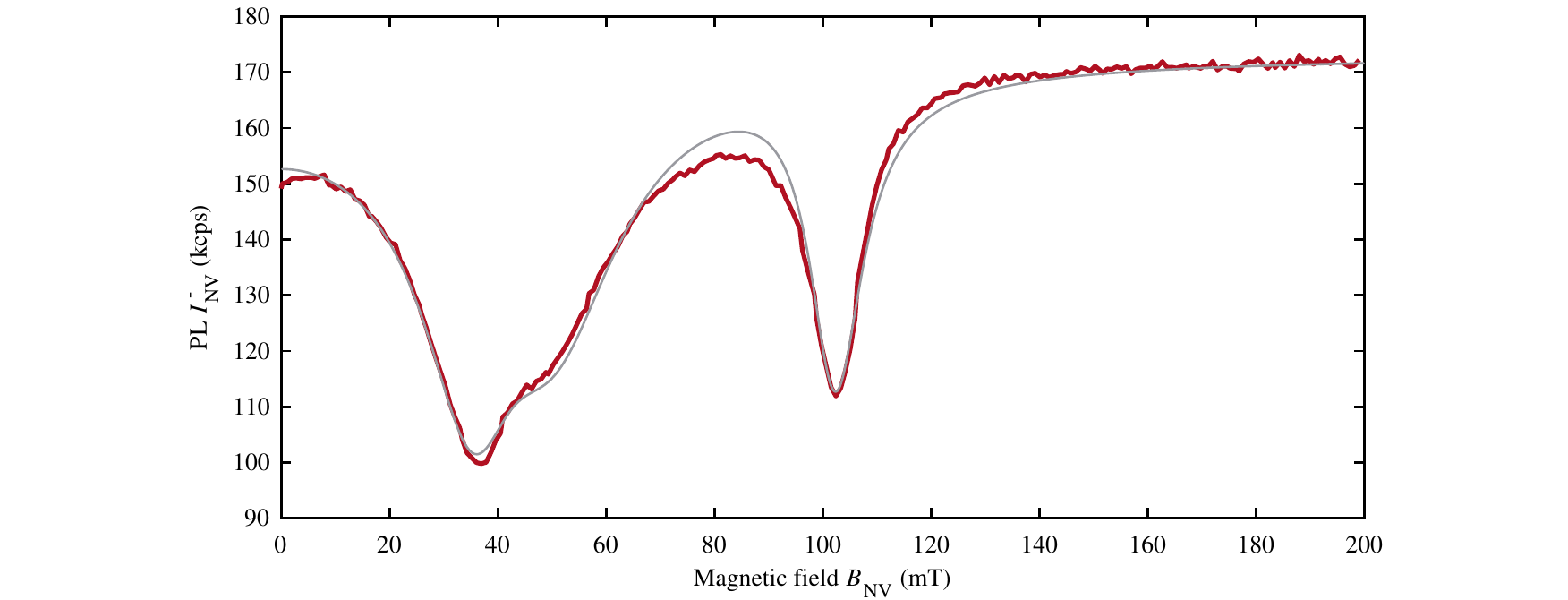}
	\caption{\NVm{} photoluminescence (PL) signal, $I_{\textrm PL}^{-}$, as a function of magnetic field, $B_{\textrm NV}$, for the NV with elevated strain.
    The fitting parameters are:
    $\delta_\perp = 77\pm2$~GHz,
    $\theta_{\delta_\perp} = 41.0\pm0.8^{\circ}$;
    B field misalignment:
    $\theta = 1.33\pm0.03^{\circ}$, $\phi = 206\pm30^{\circ}$;
    Excitation and scaling parameters:
    $\beta_{E_x} = 0.26\pm0.01$,
    $\beta_{E_y} = 0.57\pm0.04$,
    $\mathcal{I}_{0} = 26454\pm680$~cps;
	}
	\label{fig:appendix_HighStrainNV_Fitting}
\end{figure}

\begin{figure}[!ht]
	\centering
	\includegraphics[width=1\linewidth]{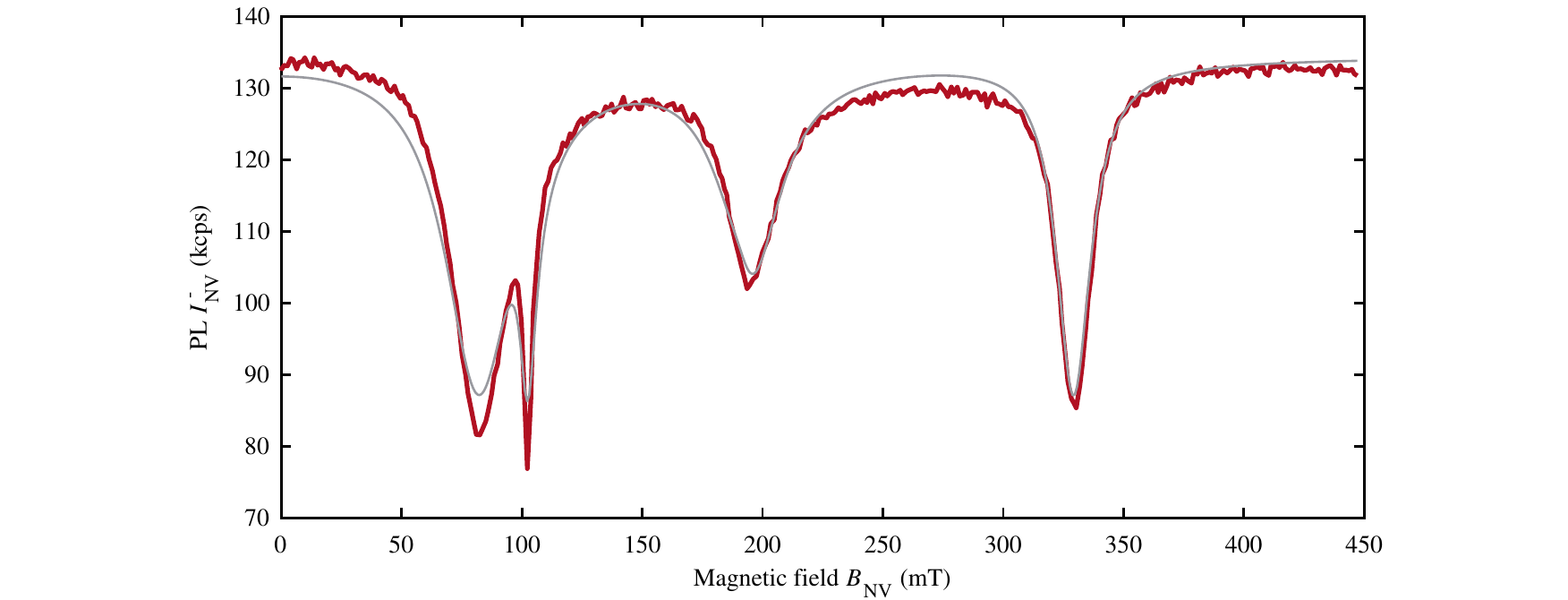}
	\caption{\NVm{} photoluminescence (PL) signal, $I_{\textrm PL}^{-}$, as a function of magnetic field, $B_{\textrm NV}$, for the low strain NV.
	The fitting parameters are:
	$\delta_\perp = 1.683\pm0.003$~GHz,
	$\theta_{\delta_\perp} = 237\pm3~^{\circ}$;
	B field misalignment:
	$\theta = 1.05\pm0.03^{\circ}$, $\phi := 0~^{\circ}$;
	Excitation and scaling parameters:
	$\beta_{E_x} = 0.296\pm0.008$,
	$\beta_{E_y} = 0.171\pm0.011$,
	$\mathcal{I}_{0} = 17927\pm765$~cps;
	}
	\label{fig:appendix_LowStrainNV_Fitting}
\end{figure}

\begin{figure}[!ht]
	\centering
	\includegraphics[width=1\linewidth]{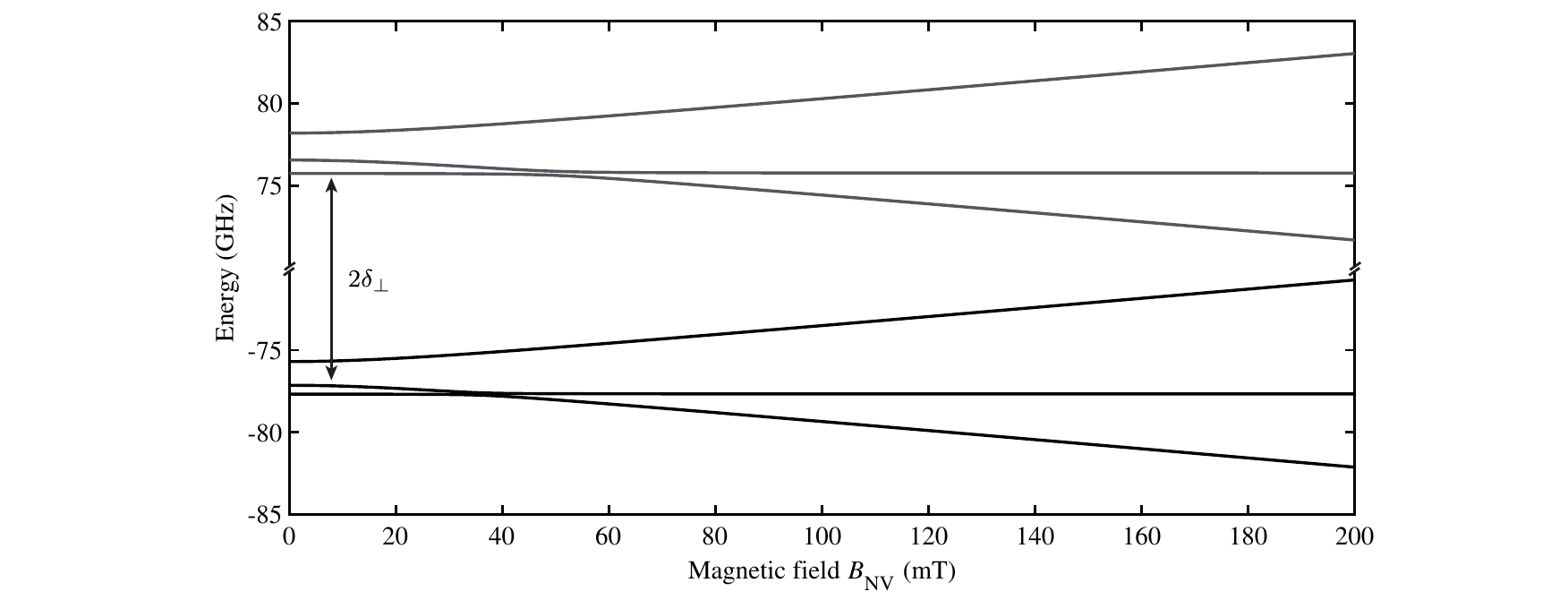}
	\caption{Energy level structure of the NV with elevated strain. The orbital branches are well separated. This is indicated by the different colors of the orbitals. 
	}
	\label{fig:appendix_HighStrainNV_Levels}
\end{figure}

\begin{figure}[!ht]
	\centering
	\includegraphics[width=1\linewidth]{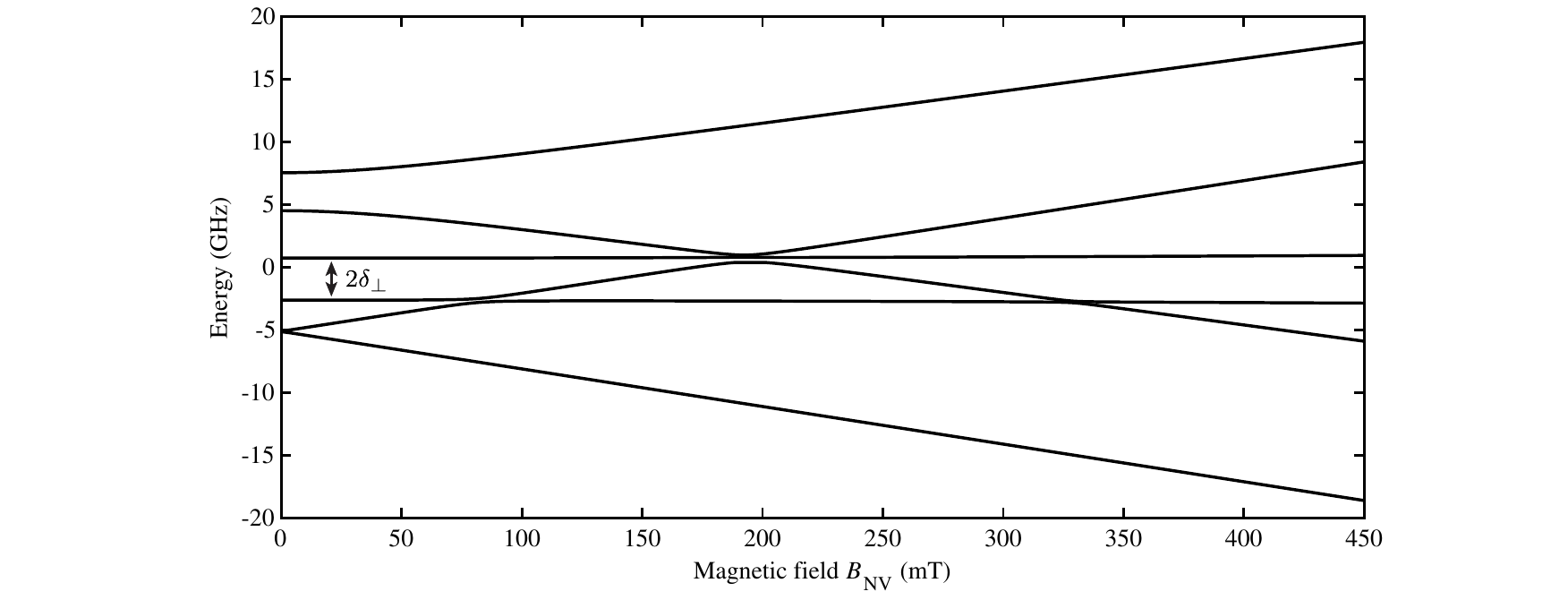}
	\caption{Energy level structure of the low strain NV. The two orbital branches are not well seperated.
	}
	\label{fig:appendix_LowStrainNV_Levels}
\end{figure}

To model and understand the NV $I_{\textrm PL}^-$ as a function of temperature and magnetic field it is important to determine intrinsic NV properties such as magnetic field alignment, strain and illumination conditions.
Using the same method as demonstrated in detail in Ref.\,\cite{Happacher2022a}, we fit the NV PL as a function of B, $I_{\textrm PL}^{-} (B)$, at the lowest obtainable temperature ($T = 2K$). 
Where the fits for both NVs is shown in Fig.~\ref{fig:appendix_HighStrainNV_Fitting} and \ref{fig:appendix_LowStrainNV_Fitting}, and the extract properties are shown in Table~\ref{table:appendix_FittingValues}.
The corresponding level diagrams for the two NVs is shown in Fig.~\ref{fig:appendix_HighStrainNV_Levels} and \ref{fig:appendix_LowStrainNV_Levels}. 
The values extracted from the $I_{\textrm PL}^-$ spectra of the two NVs at low temperature are used in the Lindbladian model discussed in Section~\ref{Subsection:Lindblad}.

\begin{table}
	\begin{tabular}{c | c | c | c | c | c | c }
		\hline
		NV~ & ~$\delta_\perp$ (GHz)~ & \quad$\phi_\delta$ ($^\circ$)~ & \quad$\theta_B$ ($^\circ$)~ & $\beta_{E_x}$  &  $\beta_{E_y}$  & ~$I_{\textrm bck}^{-}$ (kcps)~ \\
		\hline
		NV\,$\#1$ & 1.683(3) & 237(3) & 1.05(3) &  ~0.296(8)~ & ~0.171(11)~ &  17.9(8) \\
		NV\,$\#2$ & 77(2) & 41.0(8) & 1.33(3) & 0.26(1) & 0.57(4) & 26.5(7)
	\end{tabular}
\caption{Summary of the fit results for the two NVs from the $I_{\textrm{PL}}^-$ spectra at low temperature. $\eta_{\textrm{collection}}$ is fixed to 0.01. For details regarding the fitting, refer to Ref.\,\cite{Happacher2022a}}
\label{table:appendix_FittingValues}
\end{table}


\subsection{Phonon-coupling parameter}
\label{Subsection:FittingAlpha}

Using our model it is also possible to extract the phonon-coupling parameter and thus model the temperature dependence of the ESLAC structure.
Using the intrinsic NV properties that have been determined in the previous section and are shown in Table~\ref{table:appendix_FittingValues}, the temperature dependence of our model is determined by two parameters, $\alpha_{\textrm{ph}}$ and $\omega_{\textrm{c}}$
where $\alpha_{\textrm{ph}}$ is the effective coupling incorporating the phonon coupling strength to E and A phonons and $\omega_{\textrm{c}}$ is the cutoff frequency in the integral in Eq.~\ref{equation:Integral}.
These values are required to calculate the transition frequency between the two orbitals.
In our model with fix $\omega_{\textrm{c}}$ to 60~meV (See Section~\ref{Subsection:OrbitalTransitionRates}).
This leaves only one free parameter for the fit: the phonon coupling strength $\alpha_{\textrm{ph}}$.

Since the data from NV \#2 (higher strain) has a significantly better quality in laser stability over time we use this dataset to fit the model to extract the phonon-coupling strength, which gives a phonon coupling strength of $\alpha_{\textrm{ph}} = 1.70\pm0.08$~K$^5$Hz.
The parameter is determined by fitting the NV PL vs temperature across a range of magnetic field values, where an area around the ESLACs and GLSAC were excluded from the fit.
Additionally, our variable temperature insert has two temperature ranges ($T < 15$~K and $T > 15$~K), which results in different temperature gradients from the sample to the temperature sensor. 
As such for a consistent operating state for the fit we exclude the temperature range $T < 15$~K from the fit. 

For the modelling of the NV PL, we assume that the coupling strength is the same for the two NVs.
Therefore we use the determined effective coupling strength also for NV \#1.
Additionally, as both samples have been mounted in the very same way on the same PCB and measured in the same system (see~\ref{Subsection:Setup}), we also assume that there are no differences in temperature for both NVs.

The model is used to normalize the fluctuation in countrate over time as described in Section~\ref{Subsection:DataNormalization}.

Finally, we do not include additional strain effects in our model such as thermal expansion.
We think that although variations of strain on temperature should take place due to thermal expansion, considering this seems an unnecessary improvement of the model given the agreement with data.

\begin{figure}[b!]
	\centering
	\includegraphics[width=1\linewidth]{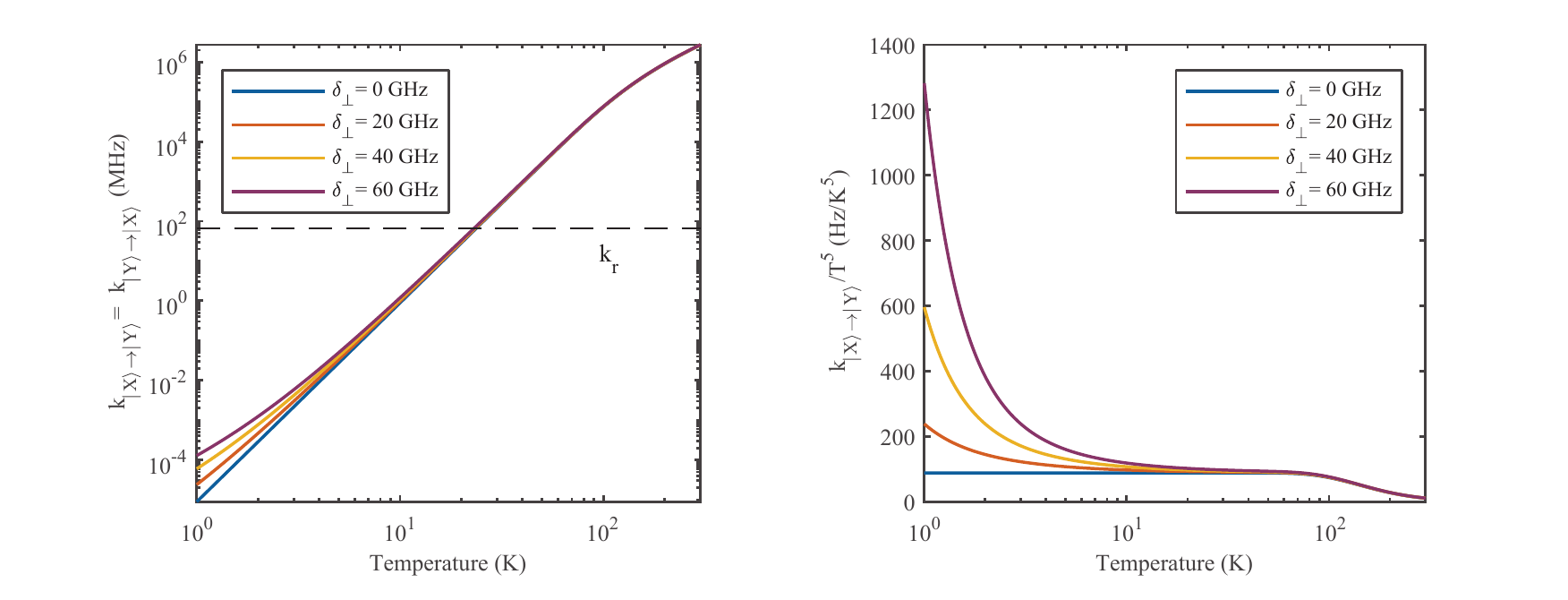}
	\caption{
       Left: Strain dependent spin-conserving transition rates between the two orbitals as a function of temperature in MHz in a double log plot. The dashed line indicates the radiative decay rate $k_r$ used in this model\,\cite{Gupta2016a}.
       Right: Strain dependent transition rates devided by temperature scaling ($T^5$).
	}
	\label{fig:appendix_MixingRate}
\end{figure}

The resulting strain-dependent spin-conserving mixing rate in the excited state orbitals is shown in Fig.\,\ref{fig:appendix_MixingRate} and shows that at low temperatures the relaxation into the ground state $k_r$ is the dominant rate.

\subsection{Comparison of the effective coupling \texorpdfstring{$\alpha_{\textrm{ph}}$}{Alpha} to literature values}\label{Subsection:comparison}

In order to compare the effective phonon coupling predicted in our model to that published in literature we need to adopt a change in the transition rate $\Gamma_{XY}$, as we also include an additional coupling of the $A_1$-symmetry phonons ($\eta_A$).
In previous work by Plakhotnik et al.~\cite{Plakhotnik2015a} the transition rate from phonon-coupling is described as 
\begin{equation}
	W = B_E T^5 I,
\end{equation}
where
\begin{equation}
	B_E = \frac{64}{\pi} \hbar \eta_E^2 k_B^5.
\end{equation}
To get a comparable form we use the defined prefactor from Eq.~\ref{Eq:phonon_coupling_prefactor}
\begin{eqnarray}
	\alpha_{\textrm{ph}} = \frac{4\pi}{\hbar} \left( \eta^2_E+\eta_A \eta_E\right) k_B^5.
\end{eqnarray}
where this term is a fit parameter in our model. 
Using the fitted value for $\alpha_{\textrm{ph}}$ and equating it to the literature $B_E$ we can solve for $\eta_E$ 
\begin{equation}
	\eta_E = \sqrt{\frac{\pi \alpha_{\textrm{ph}}}{64 \hbar k_B^5}}.
\end{equation}
Our value of $\eta_E = 163 \pm 5~(\text{MHz/meV}^3)$ extracted with this method is consistent with other measured values shown in Table~\ref{Table:EtaComp}. 

\begin{table}
	\begin{tabular}{c | c }
		Reference & $\eta_E~(\text{MHz/meV}^3)$ \\
		\hline
		Our Result & $163 \pm 5$ \\
		Plakhotnik et al.~\cite{Plakhotnik2015a} & 143 \\
		Goldmann et al.~\cite{Goldman2015a} &  $276 \pm 15$ \\
		Abtew et al.~\cite{Abtew2011a} & $196$
	\end{tabular}
\caption{Comparison of E-symmetry photon coupling $\eta_E$ from various sources.
        }
\label{Table:EtaComp}
\end{table}

\subsection{Optically detected magnetic resonance simulations} \label{Subsection: ODMR}

The model is also capable of capturing the changes or optical spin-state readout contrast (ODMR).
In order to realise this we included transition rates for microwave driving between the ground states of the NV system, e.g. $k_{MW} \neq 0$.
Then by normalising the different states with Eq.~\ref{Eq: contrast} we can extract the expected contrast.

\begin{figure}[b!]
	\centering
	\includegraphics[width=1\linewidth]{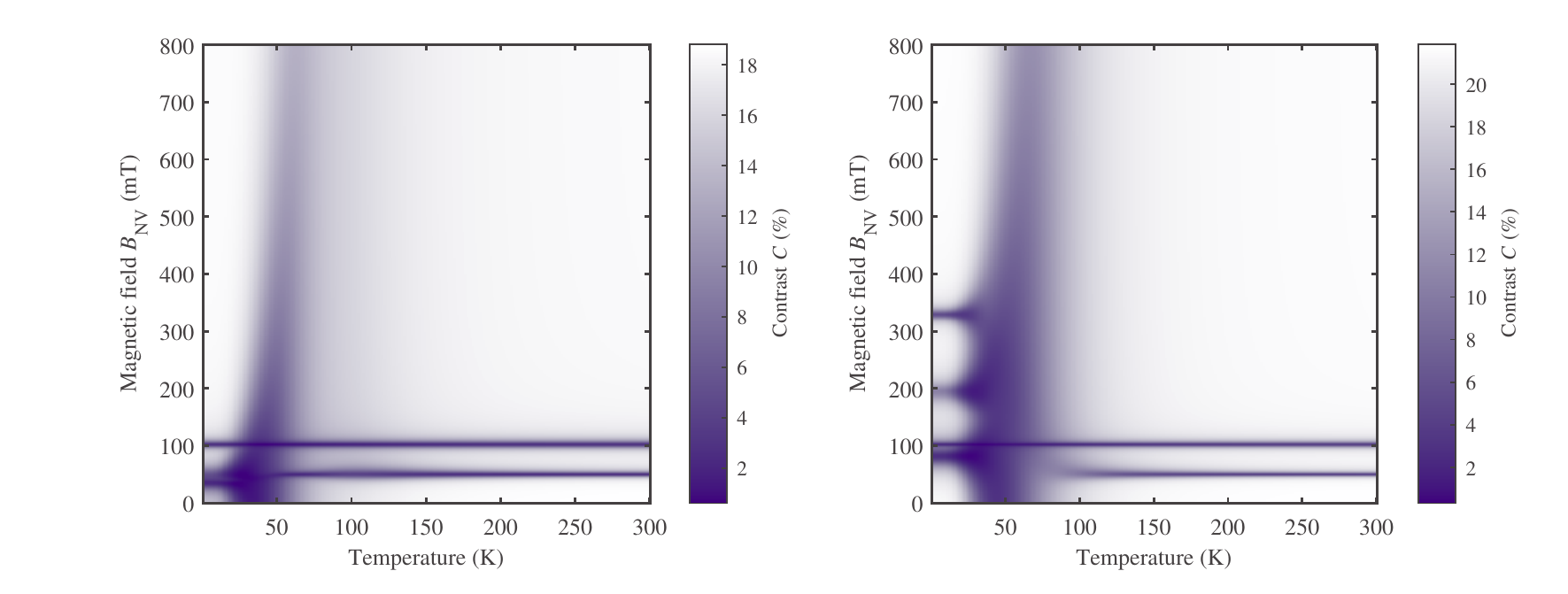}
	\caption{
     Simulated ODMR contrast $\mathcal{C}$ for the NV \#2 with the elevated strain (left) and for NV \#1 with low strain (right) for microwave driving between $\ket{2}$ (${^3}A_2^0$) and $\ket{3}$ (${^3}A_2^{-1}$) with a Rabi time of 100~ns using the parameter extracted from the fits above.
     Note that the maximum contrast is slightly different for the two NVs due to their intrinsic parameters. (See table~\ref{table:appendix_FittingValues})
	}
	\label{fig:appendix_ODMRContrastSimulation}
\end{figure}

We perform simulations of optically detected magnetic resonance (ODMR) to investigate the impact of the photoluminescence changes over temperature on the contrast.
In Fig.\,\ref{fig:appendix_ODMRContrastSimulation} we show the simulated ODMR contrast for the both NVs centers, with a $\pi$-rotation time of $\tau_\pi = 100$~ns.
As expected, the contrast follows a very similar behaviour as the PL in the simulation and data as the drop in contrast is directly related to the mixing of the states in the ESLAC.

\subsection{Emergence of the RT-ESLAC}

The emergence of the RT-ESLAC depends highly on strain as the low temperature energy configuration changes the effect orbital averaging has on the photoluminescence.
In order to quantify this, we calculate the PL contrast from the formation of the RT-ESLAC ($B = 50.5$~mT) as a function of temperature.
To isolate the effect of the RT-ESLAC, we calculate $I_{\rm PL}^- (B, T)$ where the spin-conserving mixing rate to and from the $m_s=-1$ states in the excited state are zero ($k_{\ket{X,-1}\to\ket{Y,-1}}=0$ and $k_{\ket{Y,-1}\to\ket{X,-1}}=0)$),
which we define as ${\tilde{I}}_{\rm PL}^-$. 
All the other transition rates remain the same.
This does not change the overall behaviour of the NV photophysics except that it effectively removes the RT-ESLAC from the simulation.
This can now be used as a normalisation to subtract all other effects from that simulation, which allows for the contrast due to the RT-ESLAC to be defined as
\begin{equation}
	\begin{aligned}[c]
		\mathcal{C}_{\textrm{RT-ELSAC}} = \frac{\tilde{I}^{-}_{\textrm{PL}}}{\tilde{I}^{-\,\,max}_{\textrm{PL}}} - \frac{I^{-}_{\textrm{PL}}}{I^{-\,\,max}_{\textrm{PL}}}
	\end{aligned}
\end{equation}
where the superscript max refers to the maximum PL in that simulation.
The results of this simulation are shown in Fig.~3 of the main text.

\begin{figure}[b!]
	\centering
	\includegraphics{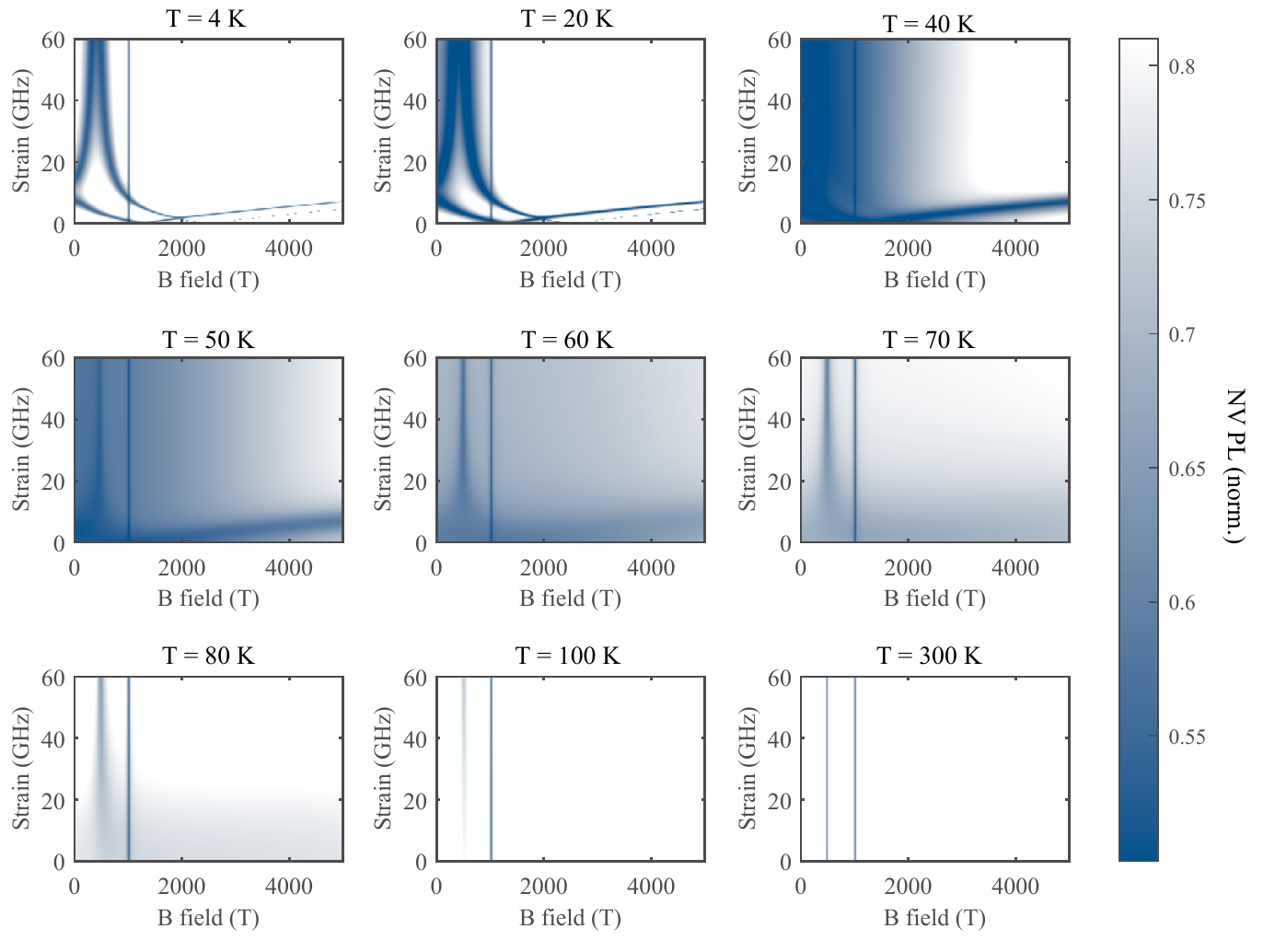}
	\caption{
    PL intensity $I_{\rm PL}^{-}$ as a function of $B_{\rm NV}$ and $\delta_\perp$, calculated with the described model for the $10$-level system illustrated. Note that the colormap has been slightly adapted to highlight the transition. 
	}
	\label{fig:appendix_TemperatureEvolution}
\end{figure}

To further illustrate this effect we show the evolution of the PL at a variety of temperatures in Fig.\,\ref{fig:appendix_TemperatureEvolution}.

%\newpage
%\bibliographystyle{apsrev4-2}
%\bibliography{Bibliography_NV_Temperature_Dependence}

\end{document}